\documentclass{JHEP3}
\usepackage{epsfig}
\def\bea{\begin{eqnarray}}
\def\eea{\end{eqnarray}}
\def\CF{{\cal F}}

\def\CN{{\cal N}}

\def\CW{{\cal W}}

\makeatletter
\def\overbracket#1{\mathop{\vbox{\ialign{##\crcr\noalign{\kern3\p@}
\downbracketfill\crcr\noalign{\kern3\p@\nointerlineskip}
$\hfil\displaystyle{#1}\hfil$\crcr}}}\limits}
\def\underbracket#1{\mathop{\vtop{\ialign{##\crcr
$\hfil\displaystyle{#1}\hfil$\crcr\noalign{\kern3\p@\nointerlineskip}
\upbracketfill\crcr\noalign{\kern3\p@}}}}\limits}
\def\overparenthesis#1{\mathop{\vbox{\ialign{##\crcr\noalign{\kern3\p@}
\downparenthfill\crcr\noalign{\kern3\p@\nointerlineskip}
$\hfil\displaystyle{#1}\hfil$\crcr}}}\limits}
\def\underparenthesis#1{\mathop{\vtop{\ialign{##\crcr
$\hfil\displaystyle{#1}\hfil$\crcr\noalign{\kern3\p@\nointerlineskip}
\upparenthfill\crcr\noalign{\kern3\p@}}}}\limits}
\def\downparenthfill{$\m@th\braceld\leaders\vrule\hfill\bracerd$}
\def\upparenthfill{$\m@th\bracelu\leaders\vrule\hfill\braceru$}
\def\upbracketfill{$\m@th\makesm@sh{\llap{\vrule\@height3\p@\@width.7\p@}}%
\leaders\vrule\@height.7\p@\hfill
\makesm@sh{\rlap{\vrule\@height3\p@\@width.7\p@}}$}
\def\downbracketfill{$\m@th
\makesm@sh{\llap{\vrule\@height.7\p@\@depth2.3\p@\@width.7\p@}}%
\leaders\vrule\@height.7\p@\hfill
\makesm@sh{\rlap{\vrule\@height.7\p@\@depth2.3\p@\@width.7\p@}}$}
\makeatother

\def\Tr{{\rm Tr}}
\def\IC{\mathbb{C}}
\newcommand{\eref}[1]{(\ref{#1})}
\newcommand{\fref}[1]{Figure~\ref{#1}}
\newcommand{\beq}{\begin{equation}}
\newcommand{\eeq}{\end{equation}}
\def\IR{\mathbb{R}}

\newcommand{\diff}[2]{\frac{\partial #1}{\partial #2}}

\preprint{UPR-1025-T, ITFA-2002-57 \\
VPI-IPPAP-02-17, {\tt hep-th/0212082}}
\title{Multi-Trace Superpotentials vs. Matrix Models}
\author{
Vijay Balasubramanian$^1$, Jan de Boer$^2$,
Bo Feng$^3$, Yang-Hui He$^1$, 
Min-xin Huang$^1$, Vishnu Jejjala$^4$, and Asad Naqvi$^1$ \\
~\\
\[
\begin{tabular}{ll}
\begin{tabular}{l}
1. David Rittenhouse Laboratories,\\
The University of Pennsylvania,\\
209  S.~33rd St.,
Philadelphia, PA 19104-6396\\
\end{tabular}
&
\begin{tabular}{l}
2. Institute of Theoretical Physics,\\
University of Amsterdam,\\
Valckenierstraat 65, 1018 XE,\\
Amsterdam, The Netherlands\\
\end{tabular}
\\
\\
\begin{tabular}{l}
3. Institute for Advanced Study, \\
Olden Lane, Princeton, NJ 08540 \\
\end{tabular}
&
\begin{tabular}{l}
4. Institute for Particle Physics and Astrophysics \\
Department of Physics, Virginia Tech \\
Blacksburg, VA 24061 \\
\end{tabular}
\end{tabular}
\]
\email{vijay@physics.upenn.edu,jdeboer@science.uva.nl,fengb@ias.edu,yanghe@physics.upenn.edu,}
\email{minxin@sas.upenn.edu,vishnu@vt.edu,naqvi@physics.upenn.edu}
}
\abstract{We consider $\CN = 1$ supersymmetric $U(N)$ field theories in
four dimensions with adjoint chiral matter and a multi-trace
tree-level superpotential.   We show that the computation of the
effective action as a function of the glueball superfield localizes to
computing matrix integrals. Unlike the single-trace case, holomorphy
and symmetries do not forbid non-planar contributions. Nevertheless,
only a special subset of the planar diagrams contributes to the exact
result. Some of the data of this subset can be
computed from the large-$N$ limit of an associated multi-trace
Matrix model.  However, the prescription {\it differs} in important
respects from that of Dijkgraaf and Vafa  for single-trace
superpotentials in that the field theory effective action is {\it not}
the derivative of a multi-trace matrix model free energy.   The basic subtlety involves the correct
identification of the field theory glueball as a variable in the Matrix model, as we show via an auxiliary construction involving a single-trace matrix model with additional singlet fields which are integrated out to compute the multi-trace results.   Along the way we also 
describe a general technique for computing the large-$N$ limits of
multi-trace Matrix models and raise the challenge of finding the field
theories whose effective actions they may compute.  Since our models
can be treated as $\CN = 1$ deformations of pure $\CN =2$ gauge theory,
we show that the effective superpotential that we compute also 
follows from the $\CN = 2$ Seiberg-Witten solution.  Finally, we observe an interesting connection
between multi-trace local theories and non-local field theory.
}
\begin{document}
%
\section{Introduction}
Dijkgraaf and Vafa have recently
made the remarkable proposal that the superpotential and other
holomorphic data of $\CN =1$ supersymmetric gauge theories in four
dimensions can be computed from an auxiliary Matrix model
\cite{DV1,DV2,DV3}.
While the original proposal arose from consideration of stringy
dualities arising in context of geometrically engineered field
theories,  
two recent papers have suggested direct field theory proofs of the proposal 
\cite{DGLVZ,CDSW}.   These works considered $U(N)$ gauge theories with
an adjoint chiral matter multiplet $\Phi$ and a tree-level
superpotential $W(\Phi) = \sum_k g_k \, \Tr(\Phi^k)$.  Using somewhat
different techniques (\cite{DGLVZ} uses properties of superspace
perturbation theory  while \cite{CDSW} relies on factorization of
chiral correlation  
functions, symmetries, and the Konishi anomaly) these papers conclude that:
\begin{enumerate}
\item The computation of the effective superpotential as a function of the
	glueball superfield reduces to computing matrix integrals. 
\item Because of holomorphy and symmetries (or properties of
	superspace perturbation theory),  
	only planar Feynman diagrams contribute.
\item These diagrams can be summed up by the large-$N$ limit of an
	auxiliary Matrix model.  The field theory effective action is
	obtained as a derivative of the Matrix model free energy. 
\end{enumerate}
Various generalizations and extensions of these ideas 
({\em e.g.}, $\CN=1^*$ theories \cite{Dorey1,Dorey2}, 
fundamental matter \cite{Argurio:2002xv,McGreevy}, quantum moduli spaces
\cite{Berenstein:2002sn}, non-supersymmetric cases
\cite{Dijkgraaf:2002wr}, other gauge groups
\cite{Ita:2002kx,niels,Ashok:2002bi,BoSO}, baryonic matter
\cite{Argurio:2002hk,Bena:2002ua}, gravitational corrections
\cite{Klemm:2002pa,Dijkgraaf:2002yn}, and Seiberg Duality
\cite{Feng:2002zb,Feng:2002yf}) 
have been considered in the recent literature. 

A stringent and simple test of the Dijkgraaf-Vafa proposal and of the
proofs presented in \cite{DGLVZ,CDSW} is to consider superpotentials
containing multi-trace terms such as 
\beq
W(\Phi) = g_2 \Tr(\Phi^2) + g_4 \Tr(\Phi^4) + \widetilde{g}_2 (\Tr(\Phi^2))^2.
\label{doubtrace}
\eeq
We will show that for such multi-trace theories: 
\begin{enumerate}
\item The computation of the effective superpotential as a function of the
glueball superfield still reduces to computing matrix integrals. 
\item  Holomorphy and symmetries do not forbid non-planar
contributions; nevertheless only a certain subset of the planar
diagrams contributes to the effective superpotential.
\item  This subclass of planar graphs also contributes to
the large-$N$ limit of an associated multi-trace Matrix model.  However, because
of differences in  
combinatorial factors, the field theory effective superpotential {\it
cannot} be obtained simply as a 
derivative of the multi-trace Matrix model free energy as in \cite{DV3}. 
\item   
Multi-trace theories can be linearized in traces by the addition of auxiliary singlet fields $A_i$.
The superpotentials for these theories as a function of both the $A_i$ and the glueball can be computed from an associated Matrix model.
This shows that the basic subtlety involves the correct
identification of the field theory glueball as a variable in a related Matrix model.
\end{enumerate}

The plan of this paper is as follows.  In Sec.~2 we carefully analyze the
methods of \cite{DGLVZ} and generalize them so that they apply
to an $\CN = 1$ $U(N)$ gauge theory in four dimensions with
a tree-level superpotential of the form  \eref{doubtrace}.   Along the way we introduce some
new techniques that deepen our understanding of the selection rules determining which perturbative field theory diagrams contribute to the effective superpotential of an $\CN = 1$ field theory.   
Using this understanding we demonstrate how the conclusions (1) and (2) above arise and show that contributing diagrams are tree-like graphs in which single-trace diagrams are pasted together by double-trace vertices through which no momentum flows.  We illustrate our results by explicitly computing the superpotential to the first few orders in perturbation theory.    Finally, we observe an intriguing connection between multi-trace local theories and non-local field theory.

Since the field content of pure $\CN = 2$ supersymmetric $U(N)$ gauge
theory in four dimensions consists in $\CN =1$ language of a 
vector multiplet $\CW_\alpha$ and an adjoint chiral multiplet
$\Phi$,  the superpotential (\ref{doubtrace}) can be treated as a
deformation of an $\CN = 2$ theory to an $\CN = 1$ theory.   Hence, we can 
use global symmetries, holomorphy, regularity conditions,
and the Seiberg-Witten solution of $\CN = 2$ gauge theory to compute
the exact superpotential.    We 
carry out this procedure in Sec.~3, using the fact that the vacuum
expectation value of the product of chiral operators
$\langle\Tr(\Phi^2)^2 \rangle$ factorizes as $\langle \Tr(\Phi^2)
\rangle^2$.
We show that the result exactly captures the subset of the
planar diagrams that
contribute to the exact field theory superpotential. 
The assumption of factorization in the Seiberg-Witten analysis is
equivalent to the  
vanishing of a certain subset of planar diagrams in our perturbative
computations. 

In Sec.~4 we demonstrate a general
technique for solving $U(M)$  matrix models (or general complex matrix
models) with multi-trace 
potentials.  The essential observation, following Das, Dhar, Sengupta,
and Wadia \cite{Das}, 
is that in the large-$M$ limit, mean field methods can be used
to solve for the effect on a single matrix eigenvalue of the rest
of the matrix.  We explain the general method and solve two examples
in detail.   The first example has a potential $V(\Phi) = M(g_2
\Tr(\Phi^2) + g_4 \Tr(\Phi^4) + {\widetilde{g}_2 \over M} \Tr(\Phi^2)^2)$
for $\phi \in U(M)$.     By expanding the exact large-$M$ result in powers of the couplings we
demonstrate how this limit computes the data relevant for a certain
subset of the planar contributions to the effective action of the
field theory with the tree-level superpotential in (\ref{doubtrace}).  In
the proposal of Dijkgraaf and Vafa \cite{DV3} and the subsequent
generalizations ({\em e.g.}, \cite{Dorey1} to \cite{Feng:2002yf}), the field
theory effective action was related simply to the free energies of
auxiliary matrix models and their derivatives.    We demonstrate the absence of such a relation between multi-trace field theories and multi-trace Matrix models.  As a  further
illustration of the mean field technique for computing large-$M$
limits, we study a matrix model with a general quartic potential.

Multi-trace field theories can be linearized in traces by the introduction of new singlet fields which can be integrated out to produce the multi-trace theory.   In Sec.~5 we show how this procedure is carried out and relate the resulting linearized superpotential to a Matrix model following the techniques of \cite{DGLVZ}.   In the Matrix model integrating out the singlets at the level of the free energy reproduces the multi-trace results that do not agree with the field theory.  However, integrating out after computing the linearized field theory superpotential leads to agreement.   This shows that the basic subtlety here involves the correct identification of the field theory glueball as a variable in an associated Matrix moedl.

It is worth mentioning several further reasons why multi-trace
superpotentials are interesting. 
First of all, the general deformation of a pure $\CN = 2$ field theory
to an $\CN = 1$ theory with adjoint matter involves multi-trace
superpotentials, and therefore these deformations are important to
understand.  What is more, multi-trace superpotentials 
cannot be geometrically engineered \cite{Reverse}
in the usual manner
for a simple reason: in geometric engineering of gauge theories
the tree-level superpotential arises from a disc diagram for open
strings on a D-brane and these, having only one boundary, produce
single-trace terms.   In this context, even if multi-trace terms
could be produced by quantum corrections, their coefficients would
be determined by the tree-level couplings and would not be freely
tunable.  Hence comparison of the low-energy physics arising from
multi-trace superpotentials with the corresponding Matrix model
calculations is a useful probe of the extent to which the 
Dijkgraaf-Vafa proposal is tied to its geometric and D-brane
origins.   In addition to these motivations, it is worth recalling that the
double  scaling limit of the $U(N)$ matrix model with a
double-trace potential is related to a theory of two-dimensional gravity
with a cosmological constant.   This matrix model also displays phase
transitions between smooth, branched polymer and intermediate phases
\cite{Das}.   It would be interesting to understand whether and how these phenomena
manifest themselves as effects  in a four dimensional field theory.
The results of our paper suggest that  these
phase transitions and the physics of two-dimensional cosmological constants are
embedded within four-dimensional field theory.  It would be interesting to
explore this.   Finally, multi-trace deformations of field
theories have recently made an appearance in the contexts of the
$AdS$/CFT correspondence and a proposed definition of string
theories with a nonlocal worldsheet theory \cite{AdS}.

%
\section{Multi-Trace Superpotentials from Perturbation Theory}
\label{derivation}

In this section we begin by reviewing the field theoretic proof that
when treated as 
a function of the glueball superfield, the effective superpotential of
an $\CN = 1$ supersymmetric gauge theory with single-trace tree-level
interactions is computed by planar matrix diagrams \cite{DGLVZ,CDSW}.
We will then describe how these arguments are modified by the presence of
multi-trace terms in the tree-level action.  Finally, we will explicitly
illustrate our reasoning by perturbatively computing the diagrams that
contribute to the effective superpotential of a multi-trace theory up to
third order in the couplings.  We will always work around a vacuum with
unbroken $U(N)$ symmetry.

\subsection{A Schematic Review of the Field Theory Superpotential Computation}

Below we give a schematic description of the methods of \cite{DGLVZ}
for the computation of the effective superpotential of an $\CN = 1$
field theory.   While \cite{DGLVZ} discussed theories with
single-trace Lagrangians, we will find that most of their arguments
will generalize easily to multi-trace theories.

\begin{description}
\newcommand\bPhi{\bar{\Phi}}
\newcommand\btheta{\bar{\theta}}
\newcommand\bD{\overline{D}}
\newcommand\bW{\overline{W}}
\newcommand\bm{\overline{m}}
\newcommand\pa{\partial}
\newcommand\vev[1]{\langle{#1}\rangle}
\item[1. The Action: ] 
The matter action for an ${\cal N}=1$ $U(N)$ gauge theory with a vector multiplet
$V$, a massive chiral superfield $\Phi$, and superpotential $W(\Phi)$,
is given in superspace by
\begin{equation}
\label{eq:action}
S(\Phi,\bPhi) = \int d^4x\, d^4\theta\, \bPhi e^V \Phi + 
\int d^4x\, d^2\theta\, W(\Phi) + \mbox{{\rm h.c.}}
\end{equation}
\item[2. The Goal: ]
We seek to compute the  effective superpotential 
as a function of the glueball superfield
\begin{equation}
S = \frac{1}{32\pi^2}\Tr(\CW^\alpha\CW_\alpha),
\end{equation}
where 
\begin{equation}
\CW_\alpha = i \bD^2 e^{-V} D_{\alpha} e^V
\end{equation} 
is the gauge field strength of $V$, with 
$D_\alpha = \pa/\pa\theta^\alpha$ and
$\bD_{\dot\alpha} = \pa/\pa\btheta^{\dot\alpha}+i\theta^\alpha\pa_{\alpha\dot\alpha}$
the superspace covariant derivatives, and 
$D^2 = \frac{1}{2} D^\alpha D_\alpha$ and
$\bD^2 = \frac{1}{2} \bD^{\dot\alpha} \bD_{\dot\alpha}$.
The gluino condensate $S$ is a commuting field constructed out of a pair
of fermionic operators $\CW_\alpha$. 
\item[3. The Power of Holomorphy: ]
We are interested in expressing the effective superpotential in terms of the
{\em chiral} glueball superfield $S$.   Holomorphy tells us that it will be independent of the parameters of the anti-holomorphic part of the tree-level superpotential.
Therefore, without loss of generality, we can choose a particularly simple form 
for $\bW(\bPhi)$:
\begin{equation}
\bW(\bPhi) = \frac{1}{2}\bm\bPhi^2.
\end{equation}
Integrating out the anti-holomorphic fields  
and performing standard superspace manipulations
as discussed in Sec.\ 2 of \cite{DGLVZ},
gives
\begin{equation}
\label{niceaction}
S = \int d^4x d^2\theta 
\left(-\frac{1}{2\bm}\Phi\left(\Box- i{\cal W}^{\alpha} D_{\alpha}\right)\Phi 
+ W_{tree}(\Phi)\right)
\end{equation}
as the part of the action that is relevant for computing the effective potential as a function of $S$.
Here, $\Box=\frac{1}{2}\pa_{\alpha\dot\alpha}\pa^{\alpha\dot\alpha}$ is the d'Alembertian,
and $W_{tree}$ is the tree-level superpotential, 
expanded as $\frac{1}{2}m\Phi^2 + \mbox{interactions}$.   (The reader may consult Sec.\ 2 of \cite{DGLVZ} for a discussion of various 
subtleties such as why the $\Box$ can be taken as the ordinary d'Alembertian as opposed to a gauge covariantized $\Box_{{\rm cov}}$).
\item[4. The Propagator: ]
After reduction into the form (\ref{niceaction}), the quadratic part gives
the propagator.  We write the covariant derivative in terms of Grassmann
momentum variables
\begin{equation}
D_\alpha = \pa/\pa\theta^\alpha := -i\pi_\alpha,
\end{equation}
and it has been shown in \cite{DGLVZ} that by rescaling the momenta we can
put $\bm=1$ since all $\bm$ dependence cancels out.  Then the momentum space representation of
the propagator is simply
\begin{equation}
\label{propagator}
\int_0^\infty ds_i\, \exp\left(-s_i(p_i^2 + {\cal W}^{\alpha} \pi_{i\alpha} + m) \right),
\label{schwinger}
\end{equation}
where $s_i$ is the Schwinger time parameter of $i$-th Feynman propagator. 
Here the precise form of the $\CW^\alpha \pi_\alpha$ depends on the representation of the gauge group that is carried by the field propagating in the loop.  
\item[5. Calculation of Feynman Diagrams: ]
The effective superpotential as a function of the glueball $S$ is  a sum of vacuum Feynman diagrams computed in the background of a fixed constant $\CW_\alpha$ leading to insertions of this field along propagators.  In general there will be $\ell$ momentum loops, and the corresponding momenta
must be integrated over yielding the contribution 
\begin{eqnarray}
I &=& \left(\int\prod_{a,i} d^4p_a\, ds_i\, e^{-s_i p_i^2}\right) \cdot
\left(\int\prod_{a,i} d^2\pi_a\, ds_i\, e^{-s_i\CW^\alpha\pi_{i\alpha}}\right) \cdot
\left(\int\prod_i ds_i e^{-s_i m}\right) \nonumber \\
&=& I_{boson} \cdot I_{fermion} \cdot \frac{1}{m^P}
\label{mominteg}
\end{eqnarray}
to the overall amplitude. 
Here $a$ labels momentum loops, while  $i = 1, \ldots, P$ labels propagators.  The momenta in the propagators are linear combinations of the loop momenta because of momentum conservation.
\item[6. Bosonic Momentum Integrations: ] 
The bosonic contribution can be expressed as 
\begin{equation}
I_{boson} = \int\prod_{a=1}^\ell \frac{d^4p_a}{(2\pi)^4} 
            \exp\left[-\sum_{a,b} p_a M_{ab}(s) p_b\right] 
          = \frac{1}{(4\pi)^{2\ell}} \frac{1}{(\det\, M(s))^2},
\end{equation}
where we have defined the momentum of the $i$-th propagator 
in terms of the independent loop momenta $p_a$
\begin{equation}
p_i = \sum_{a} L_{ia} p_a
\end{equation}
via the matrix elements $L_{ia}\in \{0,\pm 1\}$ 
and
\begin{equation}
M_{ab}(s) = \sum_i s_i L_{ia} L_{ib}.
\end{equation}
\item[7. Which Diagrams Contribute: ] 
Since each momentum loop comes with two fermionic $\pi_\alpha$ integrations (\ref{mominteg}) a non-zero amplitude will require the insertion of $2\ell$ $\pi_\alpha$s.  From (\ref{schwinger}) we see that 
that $\pi_\alpha$ insertions arise from the power series expansion of the fermionic part of the 
propagator and that each $\pi_\alpha$ is accompanied by a $\CW_\alpha$.  So in total we expect an
amplitude containing $2\ell$ factors of $\CW_\alpha$.
Furthermore, since we wish to compute the superpotential as a function of $S \sim \Tr(\CW^\alpha \CW_\alpha)$ each index loop can only have 
zero or two $\CW_\alpha$ insertions. These considerations together
imply that if a diagram contributes to the effective superpotential as function of the $S$, then number of index loops $h$ must be greater than or equal to the number of momentum loops $\ell$, {\em i.e.},
\begin{equation}
h \geq \ell.
\label{hbigl}
\end{equation}

\item[8. Planarity: ]
The above considerations are completely general. Now let us specialize
to $U(N)$ theories with single-trace operators. A diagram with $\ell$
momentum loops has 
\begin{equation}
h = \ell + 1 - 2g 
\label{eq:loops}
\end{equation}
index loops, where $g$ is the genus of the surface generated by 't Hooft double line notation.  Combining this with (\ref{hbigl}) tell us that $g  = 0$, {\em i.e.}, only planar diagrams contribute.

\item[9. Doing The Fermionic Integrations: ]
First let us discuss the combinatorial factors that arise from the
fermionic integrations.   Since the number of momentum loops is one
less than the number of index loops, we must choose which of the latter
to leave free of $\CW_\alpha$ insertions.  This gives a combinatorial
factor of $h$, and the empty index loop gives a factor of $N$
from the sum over color.   For
each loop with two $\CW_\alpha$ insertions we  
get a factor of ${1 \over 2} \CW^\alpha \CW_\alpha = 16 \pi^2 S$.
Since we are dealing with adjoint matter, the action of
$\CW_\alpha$ is through a commutator
\begin{equation}
\exp\left(-s_i[\CW_i^\alpha,-]\pi_{i\alpha}\right)
\label{commprop}
\end{equation}
in the Schwinger term. (See the appendix of \cite{Ita:2002kx} for a
nice explanation of this notation as it appears in \cite{DGLVZ}.   In
Sec.\ 2.2 we will give an alternative discussion of the fermionic
integrations that clarifies various points.) 
As in the bosonic integrals above, it is convenient to express the
fermionic propagator momenta as sums of the independent loop momenta: 
\begin{equation}
\pi_{i\alpha} = \sum_a L_{ia} \pi_{a\alpha},
\end{equation}
where the $L_{ia}$ are the same matrix elements as introduced above.
The authors of \cite{DGLVZ} also find it convenient to introduce
auxiliary fermionic variables via the equation 
\begin{equation}
\CW_i^\alpha = \sum_a L_{ia} \CW_a^\alpha \, .
\label{auxferm}
\end{equation}
Here, the $L_{ia}=\pm 1$ denotes the left- or right-action of the commutator. 
In terms of the $\CW_a^\alpha$, the fermionic contribution to the
amplitude  can be written as 
\begin{eqnarray}
I_{fermion} &=& N h(16\pi^2 S)^\ell \int \prod_a d^2\pi_a\, d^2\CW_a\,
\exp\left[-\sum_{a,b} \CW_a^\alpha M_{ab}(s) \pi_{b\alpha}\right] \nonumber \\
&=& (4\pi)^{2\ell} N h S^\ell (\det\, M(s))^2.
\label{planferm}
\end{eqnarray}

\item[10. Localization: ] 
The Schwinger parameter dependence in the bosonic and fermionic momentum integrations cancel exactly 
\begin{equation}
\label{localise}
I_{boson} \cdot I_{fermion} = N h S^\ell,
\label{prefact}
\end{equation}
implying that the computation of the effective superpotential as a
function of the $S$ localizes to summing matrix integrals.   All the
four-dimensional spacetime dependence has washed out. 
The full effective superpotential $W_{eff}(S)$ is thus a sum over
planar matrix  
graphs with the addition of the Veneziano-Yankielowicz term for 
the pure Yang-Mills theory \cite{veneziano}.  The terms in the
effective action proportional to $S^\ell$ arise exclusively from
planar graphs with $\ell$ momentum loops giving a perturbative
computation of the {\it exact} superpotential. 

%

%
\item[11. The Matrix Model: ]
The localization of the field theory computation to a set of planar matrix diagrams suggests that the sum of diagrams can be computed exactly by the large-$M$ limit of a bosonic Matrix model.  (We distinguish between $M$, the rank of the matrices in the Matrix model and $N$, the rank of the gauge group.)  The prescription of Dijkgraaf and Vafa does exactly this for single-trace superpotentials.   Since the number of momentum loops is one less than the number of index loops in a planar diagram, the net result of the bosonic and fermionic integrations in (\ref{prefact}) can be written as 
\begin{equation}
    I_{boson} \cdot I_{fermion}  = N {\partial S^h \over \partial S}.
\end{equation}
Because of this, the perturbative part of the effective superpotential, namely the sum over
planar diagrams in the field theory, can be written in terms of the genus
zero free energy $\CF_0(S)$ of the corresponding matrix model:
\begin{eqnarray}
W_{pert}(S) &=&  N \frac{\partial}{\partial S} \CF_0(S),  \label{deriv} \\
\CF_0(S) &=& \sum_h \CF_{0,h} \, S^h.
\end{eqnarray}
This free energy is conveniently isolated by taking the large-$M$ limit of the zero-dimensional
one-matrix model with $M\times M$ matrices\footnote{The original papers of
Dijkgraaf and Vafa \cite{DV1, DV2} consider $M\times M$ Hermitian matrices
({\em i.e.} matrices with real eigenvalues $\lambda_i$).  In fact, we
should think of the matrices as belonging to $GL(M,\IC)$ with eigenvalues
distributed along contours in the complex plane rather than along domains
on the real axis.  The prior results do not depend crucially on this point.
Indeed, they carry through exactly by analytic continuation.  We thank
David Berenstein for emphasizing this to us.} $\Phi$ and potential
$W(\Phi)$ whose partition function is given by
\begin{equation}
Z = \exp(M^2 \CF_0) = 
\frac{1}{{\rm Vol}(U(M))} \int [D\Phi] \exp\left(-\frac{1}{g_s}\Tr\, W(\Phi)\right) .
\label{matmod}
\end{equation} 
In this matrix model every index loop gives a power of $M$ just as in the field theory computation, and all but one index loop gives a power of $S$.  Because of this simple fact the powers of the gluino condensate  in the field theory superpotential can be conveniently counted by identifying it with the 't Hooft coupling $S \equiv M g_s$, and then differentiating the matrix model free energy as in (\ref{deriv}).    Rather surprisingly the Veneziano-Yankielowicz term  in $W_{eff}(S)$ arises from the volume factor in the integration over matrices in (\ref{matmod}).

\end{description}

One important unanswered question is why the low-energy dynamics simplifies so much when written in terms of the gluino condensates.

\subsection{Computation of a Multi-Trace Superpotential}
\begin{figure}
  \begin{center}
 \epsfysize=3.0in
   \mbox{\epsfbox{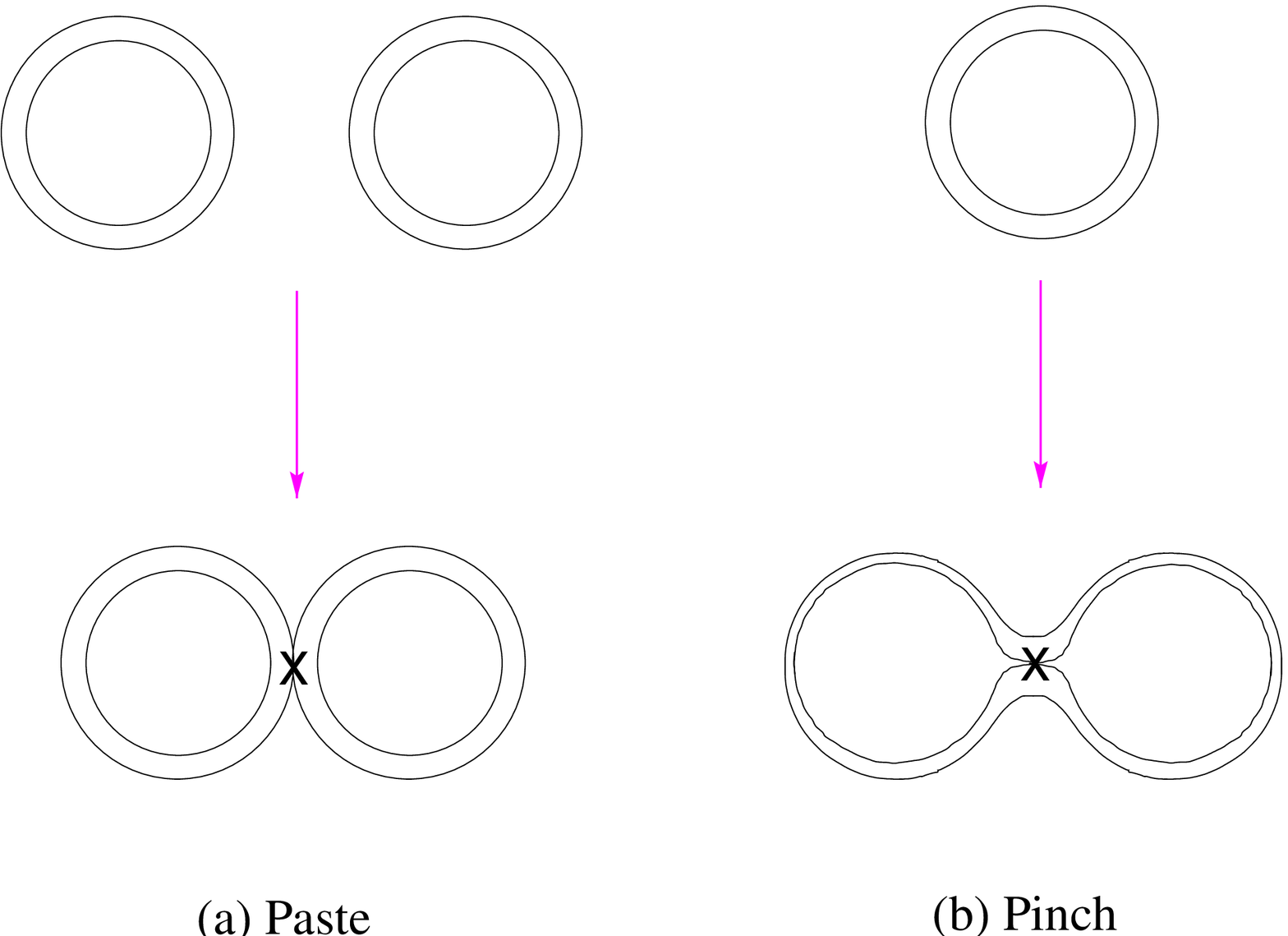}}
\end{center}
\caption{With the inclusion of the double-trace term we need new
types of vertices. These can be obtained from the ``primitive''
diagrams associated with the pure single-trace superpotential by (a)
pasting or (b) pinching. The vertices have been marked with a cross.} 
\label{pastepinch}
\end{figure}

We have reviewed above how 
the field theory calculation of the effective superpotential for a
single-trace theory localizes to a matrix model computation.  In this
subsection we show how the argument is modified when the tree-level
superpotential includes multi-trace terms.    
We consider an $\CN=1$ theory with the tree-level superpotential
\begin{equation}
\label{tree}
W_{tree}=\frac{1}{2} \Tr(\Phi^2)+g_4\Tr(\Phi^4)+\widetilde {g}_2(\Tr(\Phi^2))^2 \ .
\end{equation}
To set the stage for our perturbative computation of the effective
superpotential we begin by analyzing the structure of the new diagrams
introduced by the double-trace term.  If $\widetilde{g}_2 = 0$, the
connected diagrams we get are the familiar single-trace ones; we will
call these {\it primitive diagrams}. 
When $\widetilde{g}_2 \neq 0$  propagators in primitive diagrams can be
spliced together by new double-trace vertices. It is useful to do an
explicit example to see how this splicing occurs. 
\EPSFIGURE[h]{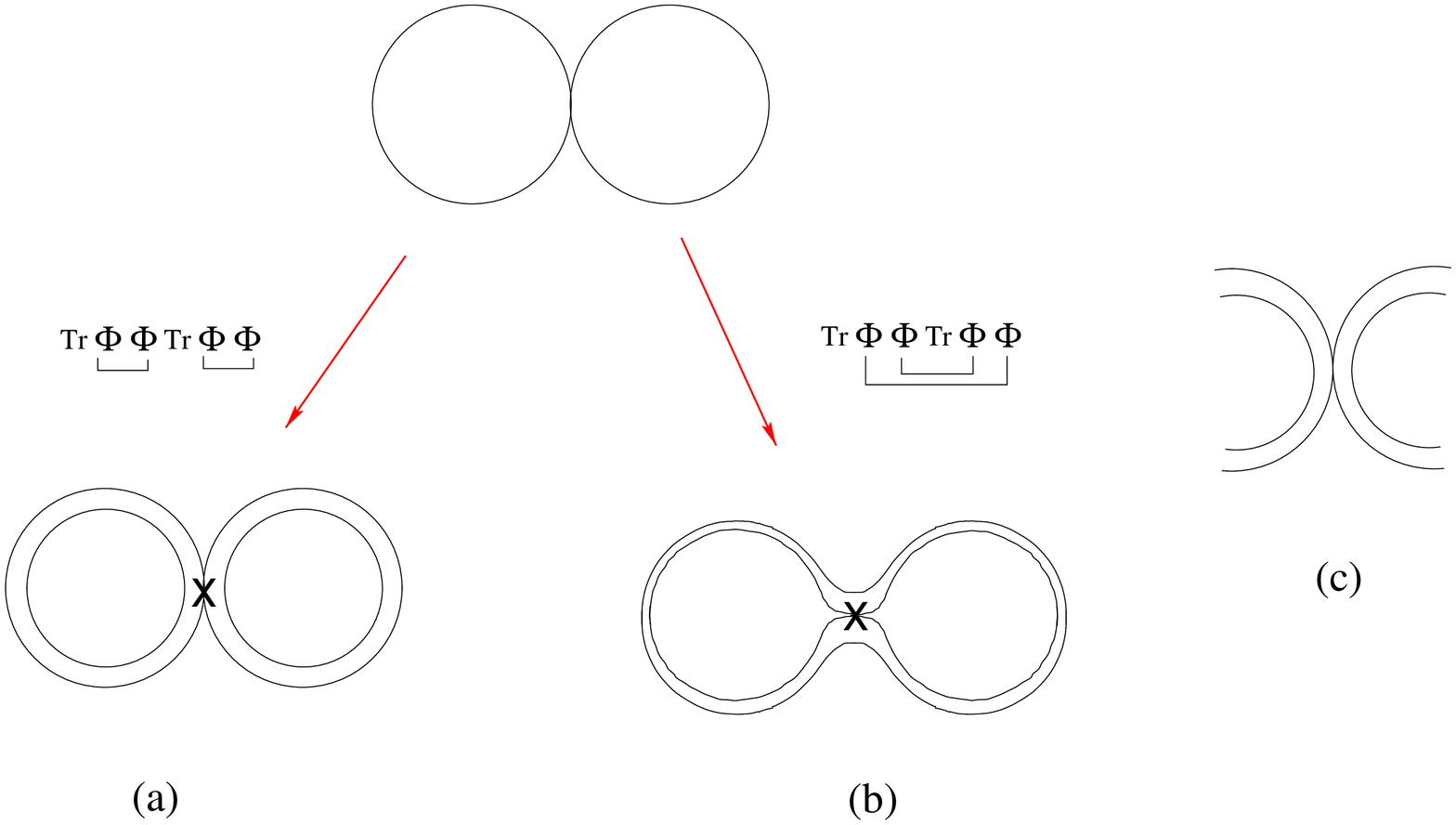,width=13cm}
{Two ways in which the double-trace operator:
$\Tr(\Phi^2) \Tr(\Phi^2)$ can be contracted using the vertex shown in
(c).
\label{f:contract}
}
As an example, let us study the expectation value of the double-trace operator:
$\langle \Tr(\Phi^2) \Tr(\Phi^2) \rangle$. 
To lowest order in couplings, the two ways to contract $\Phi$s give
rise to the two diagrams in \fref{f:contract}.  When  
we draw these diagrams in double line notation, we find that
\fref{f:contract}a  corresponding to 
 $\Tr (\overbracket{\Phi \Phi}) \Tr (\overbracket{\Phi \Phi})$ has
four index loops,  
while \fref{f:contract}b 
corresponding to
$\Tr (\overbracket{\Phi \overbracket{\Phi) \Tr (\Phi} \Phi})$
has only two index loops.   Both these graphs have two momentum loops.
For our purposes both of these Feynman diagrams can also be generated
by a simple pictorial algorithm:  we splice together propagators of
primitive diagrams using the vertex in \fref{f:contract}c, as
displayed in \fref{pastepinch}a and b.   All graphs of the
double-trace theory can be generated from primitive diagrams by this
simple algorithm.  Note that the number of index loops never changes
when primitive diagrams are spliced by this pictorial algorithm.

\begin{figure}
  \begin{center}
 \epsfysize=1.5in
   \mbox{\epsfbox{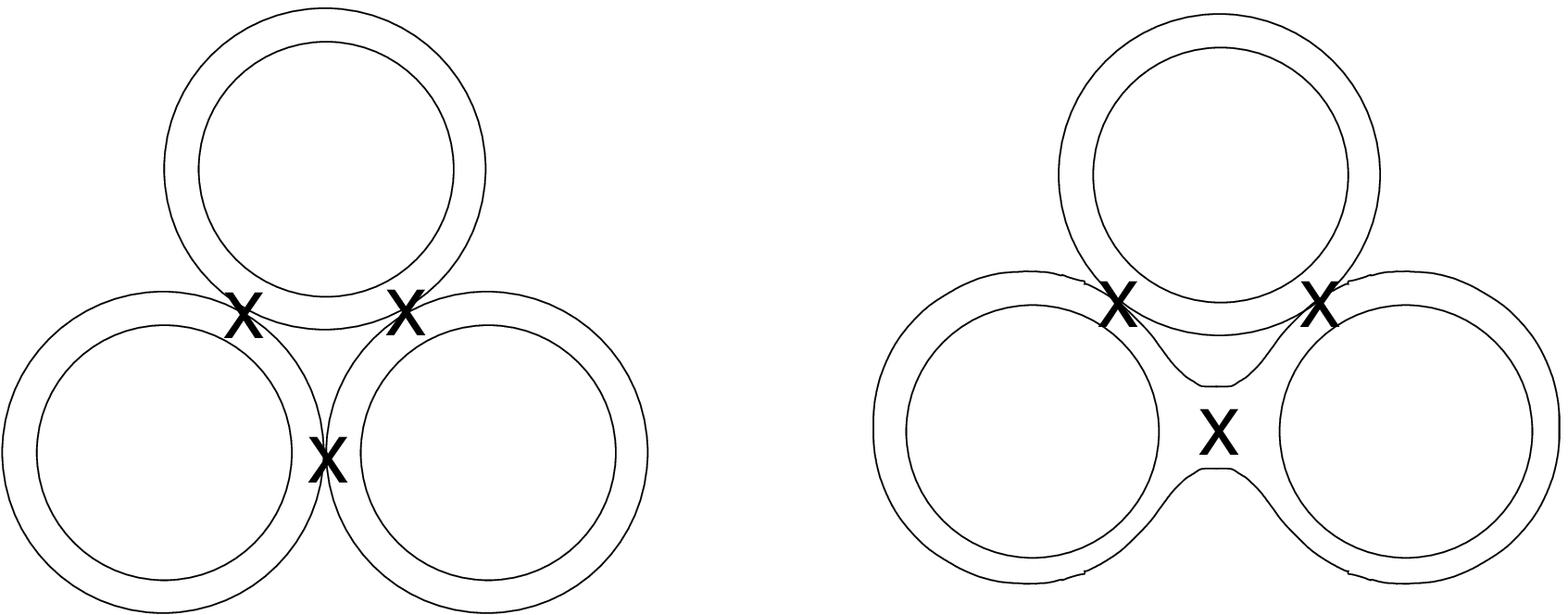}}
\end{center}
\caption{More examples of ``pinched'' diagrams.}
\label{notpaste}
\end{figure}

If a splicing of diagrams does not create a new momentum loop we say
that the diagrams have been {\it pasted} together.   This happens when
the diagrams being spliced are originally disconnected as, for
example, in \fref{pastepinch}a.  In fact because of momentum
conservation, no momentum at all flows between pasted diagrams.  If a
new momentum loop is created we say that that the diagrams have been
{\it pinched}.
This happens when two propagators within an already connected diagram
are spliced together as, for example, in \fref{pastepinch}b.   In this
example one momentum loop becomes two because momentum can flow
through the double-trace vertex.  Further examples of pinched diagrams
are given in \fref{notpaste} where the 
new loop arises from momentum flowing between the primitive diagrams
via double-trace vertices. 

To make the above statement more clear, let us provide some
calculations. First, according to our operation, the number of 
double index loops
never increases whether under pasting or pinching.
Second, we can calculate the total number of independent
momentum loops $\ell$ by
$\ell= P-V+1$ where $P$ is the number of propagators and $V$,
the number of vertices. If we connect two separate diagrams by 
{\it pasting}, 
we will have $P_{tot}=(P_1-1)+(P_2-1)+4$, $V_{tot}=V_1+V_2+1$ and 
\beq
\ell_{tot}=P_{tot}-V_{tot}+1=\ell_1+\ell_2,
\eeq 
which means that the total number of momentum 
loops is just the sum of the
individual ones.
If we insert the double-trace vertex in a single 
connected diagram by {\it pinching},
we will have $P_{tot}=P-2+4$, $V_{tot}=V+1$, and 
\beq
\ell_{tot}= P_{tot}-V_{tot}+1 = \ell+1,
\eeq
which indicates the creation of one new momentum loop.

Having understood the structure of double-trace diagrams in this way,
we can adapt the techniques of \cite{DGLVZ} to our case.  The steps
1-6 as described in Sec.\ 2.1 go through without modification since
they are independent of the details of the tree-level superpotential.
However the steps 7-11 are modified in various ways.   First of all
naive counting of  powers of fermionic momenta as in step 7 leads 
to the selection rule 
\begin{equation}
  h \geq \ell,
\label{selectno}
\end{equation}
where $h$ is the total number of index loops and $\ell$ is the total
 number of momentum loops.  (The holomorphy and symmetry based
 arguments of \cite{CDSW} would lead to the same conclusion.)   
 Since 
no momentum flows between pasted primitive diagrams it is clear that
 this selection rule would permit some of the primitive components to
 be non-planar.   Likewise, both planar and some non-planar pinching
 diagrams are admitted.   An example of a planar pinching diagram that
 can contribute according to this rule is \fref{pastepinch}b.
 However, we will show in the next subsection that more careful
 consideration of the structure of perturbative diagrams shows that
 only diagrams built by pasting planar primitive graphs give non-zero
 contributions to the effective superpotential.

\subsection{Which Diagrams Contribute: Selection Rules}

In order to explain which diagrams give non-zero contributions to the
multi-trace superpotential it is useful to first give another
perspective on the fermionic momentum integrations described in steps
7--9 above.   A key step in the argument of \cite{DGLVZ} was to split
the glueball insertions up in terms of auxiliary fermionic variables
associated with each of the momentum loops as in (\ref{auxferm}).   We
will take a somewhat different approach. 
In the end we want to attach zero or two fields ${\cal
W}^{\alpha}_{(p)}$ to each {\it index} loop, where $p$ labels the
index loop, and the  
total number of such fields must bring down enough fermionic momenta
to soak up the corresponding integrations.   On each oriented  
propagator, with momentum $\pi_{i\alpha}$, we have a left index
line which we  label $p_L$ and a right index line which we label $p_R$.
Because of the commutator in (\ref{commprop}), the contribution of
this propagator will be
\beq 
\label{e1}
\exp( - s_i ( \pi_{i\alpha} ({\cal W}^{\alpha}_{(p_L)} - 
 {\cal W}^{\alpha}_{(p_R)} )).
\label{indexW}
\eeq
Notice that we are omitting $U(N)$ indices, which are
simply replaced by the different index loop labels.  In a standard
 planar diagram for  
a single-trace theory, we have one more index loop than momentum loop.
 So even in this case the choice of auxiliary variables in (\ref{e1})
 is not quite the same as in (\ref{auxferm}), since the number of
 $\CW_\alpha$s is twice the number of index loops in (\ref{e1}) while
the number of auxiliary variables is
twice the number of momentum loops in (\ref{auxferm}).

Now in order to soak up the fermionic $\pi$  integrations in
(\ref{mominteg}), we must expand (\ref{e1}) in powers and  extract 
terms of the form 
\beq
{\cal W}_{(p_1)}^2 {\cal W}_{(p_2)}^2 \ldots {\cal W}_{(p_l)}^2,
\eeq
where $\ell$ is the number of momentum loops and all the $p_i$ are
distinct. The range of $p$ is over $1, \ldots, h$, with $h$ the
number of index loops. In the integral over the anticommuting momenta,
we have all 
$h$ ${\cal W}_{(p)}$ appearing. However, one linear combination,
which is the `center of mass' of the ${\cal W}_{(p)}$, does not
appear. This can be seen from (\ref{e1}): if we add a constant 
to all ${\cal W}_{(p)}$ simultaneously, the propagators do not
change. Thus, without loss of generality, one can set the 
${\cal W}_{(p)}$ corresponding to the outer loop in a planar diagram 
equal to zero.
Let us assume this variable is  ${\cal W}_{(h)}$
and later reinstate it. All ${\cal W}_{(p)}$ corresponding to inner
index loops remain, leaving as many of these as there are momentum
loops in  a planar diagram. It is then straightforward to demonstrate
that the $\CW$ appearing in (\ref{auxferm}) in linear combinations
reproduce the relations between propagator momenta and loop
momenta.   In other words, in this ``gauge" where the $\CW$
corresponding to the outer loop is zero, we recover the decomposition
of  
$\CW_\alpha$ in terms of auxiliary fermions associated to momentum
loops that was used in \cite{DGLVZ} and reviewed in (\ref{auxferm})
above. 

We can  now reproduce the overall factors arising from the fermionic
integrations in the planar diagrams contributing to (\ref{planferm}).
The result from the $\pi$ integrations is some constant times
\beq
\prod_{p=1}^\ell {\cal W}_{(p)}^2.
\label{res1}
\eeq
Reinstating ${\cal W}_{(h)}$ by undoing the gauge choice, namely by shifting
\beq
{\cal W}_{(p)} \rightarrow {\cal W}_{(p)} + {\cal W}_{(h)}
\eeq
for $p=1,\ldots, h-1$, (\ref{res1}) becomes
\beq
\prod_{p=1}^\ell ({\cal W}_{(p)} + {\cal W}_{(h)})^2.
\eeq
The terms on which each index loop there has either zero or
two ${\cal W}$ insertions are easily extracted:
\beq
\sum_{k=1}^h \left(\prod_{p\neq k} {\cal W}_{(p)}^2\right).
\eeq
In this final result we should replace each of the ${\cal
W}_{(p)}^2$ by $S$, and therefore the final result is of the
form
\beq
h S^{h-1},
\eeq
as derived in \cite{DGLVZ} and reproduced in (\ref{planferm}).

Having reproduced the result for single-trace theories we can easily
show that all non-planar and pinched contributions to the multi-trace
effective superpotential vanish.   Consider any diagram with $\ell$
momentum loops and $h$ index loops. 
By the
same arguments as above, we attach some ${\cal W}_{(p)}$ to each
index loop as in (\ref{indexW}), and again, the `center of mass'
decouples due to the 
commutator nature of the propagator. Therefore, in the momentum 
integrals, only $h-1$ inequivalent ${\cal W}_{(p)}$ appear.
By doing $\ell$ momentum integrals, we generate a polynomial
of order $2\ell$ in the $h-1$ inequivalent ${\cal
W}^{\alpha}_{(p)}$. This polynomial can by Fermi statistics only
be non-zero if $\ell \leq h-1$: ${\cal W}_{(p)}^3$ is zero for all
$p$. Therefore, we reach the important conclusion that  the total
number of index loops must be larger 
than the number of momentum loops 
\begin{equation}
h > \ell
\end{equation}
while the naive selection rule (\ref{selectno}) says that it
could be larger or equal.

Consider pasting and pinching $k$ primitive diagrams together, each
with $h_i$ index loops and $\ell_i$ momentum loops.  According to the
rules set out in the previous subsection, the total number of index
loops and the total number of momentum loops are given by: 
\begin{equation}
h = \sum_i h_i   ~~~;~~~  \ell \geq \sum_i \ell_i
\end{equation}
with equality only when all the primitive diagrams are pasted together
without additional momentum loops.  Now the  
total number of independent $\CW$s that appear in full diagram is
$\sum_i (h_i - 1)$ since in each primitive diagram the ``center of
mass'' $\CW$ will not appear.  So the full diagram is non-vanishing  
only when 
\begin{equation}
\ell \leq \sum_i (h_i -1).
\end{equation}
This inequality is already saturated by the momenta appearing in the
primitive diagrams if they are planar.  So we can conclude two things.
First, only planar primitive diagrams appear in the full diagram.
Second, only pasted diagrams are non-vanishing, since pinching
introduces additional momentum loops which would violate this
inequality.

\paragraph{Summary: } The only diagrams that contribute to the
effective multi-trace superpotential are pastings of planar primitive
diagrams.  These are tree-like diagrams which string together
double-trace vertices with ``propagators'' and ``external legs'' which
are themselves primitive diagrams of the single-trace theory.   Below
we will explicitly evaluate such diagrams and raise the question of
whether there is a generating functional for them.

\subsection{Summing Pasted Diagrams}

In the previous section we generalized steps 7 and 8 of the the single
trace case in Sec.\ 2.1 to the double-trace theory, and found that the
surviving diagrams consist of planar connected primitive vacuum graphs
pasted together with double-trace vertices.   Because of momentum
conservation, no momentum can flow through the double-trace vertices
in such graphs.  Consequently the fermionic integrations and the proof of
localization can be carried out separately for each primitive grapg,
and the entire diagram evaluates to a product of the the primitive
components times a suitable power of $\widetilde{g}_2$, the double-trace
coupling.  

Let $G_i$, $i = 1, \ldots, k$ be the planar primitive graphs that have
been pasted together, each with $h_i$ index loops and $\ell_i = h_i - 1$
momentum loops to make a double-trace diagram $G$.  Then, using the
result (\ref{prefact}) for the single-trace case, 
the Schwinger parameters in the bosonic and fermionic momentum
integrations cancel giving a factor 
\begin{equation}
I_{boson} \cdot I_{fermion} = \prod_i (N h_i S^\ell_i) = N^k S^{\sum_i
(h_i - 1) } \prod_i h_i,
\end{equation}
where the last factor arises from the number of ways in which the
glueballs $S$ can be inserted into the propagators of each primitive
diagram.  Defining $C(G) = \prod_i h_i$ as the glueball symmetry
factor, $k(G)$ as the number of primitive components, $h(G) = \sum_i
h_i$ as the total number of index loops and  $\ell(G) = \sum_i \ell_i =
h(G) - k(G)$ as the total number of momentum loops, we get 
\begin{equation}
I_{boson} \cdot I_{fermion} = \prod_i (N h_i S^\ell_i) = N^{h(G) - \ell(G)} S^{l(G)} C(G).
\end{equation}
We can assemble this with the Veneziano-Yankielowicz contribution
contribution for pure gauge theory \cite{veneziano} to write the
complete glueball effective action as
\begin{equation}
W_{eff}=-NS(\log (S/\Lambda^2)-1)+\sum_{G} C(G){\cal
F}(G)N^{h(G)-\ell(G)}S^{\ell(G)},
\label{effact}
\end{equation}
where $\CF (G)$ is the combinatorial factor for generating the graph
$G$ from the Feynman diagrams of the double-trace theory. 
Notice that in our
discussion, we have set $g_2=m=1$, so $\Lambda^2$ in this
equation is in fact $m \Lambda^2$ which matches the dimension
of $S$. We can
define a free energy related to above diagrams as
\beq \label{free0}
{\cal F}_0= \sum_{G}{\cal F}(G)S^{h(G)}.
\eeq
${\cal F}_0$ is a generating function for the diagrams that contribute to the effective superpotential, but does not include the combinatorial factors arising from the glueball insertions.   In the single-trace case that combinatorial factor was simply $N h(G)$ and so we could write $W_{eff} = N (\partial {\cal F}_0/\partial S)$.   Here $C(G) = \prod h_i$ is a product rather than a 
sum $h(G) = \sum h_i$, and so the effective superpotential cannot be written as a derivative of the free energy.

Notice that if we rescale $\widetilde{g}_2$
to $\widetilde{g}_2/N$, there will be a $N^{-(k(G)-1)}$ factor from 
 $k(G)-1$ insertions of the double-trace vertex. This factor will
change the $N^{h(G)-l(G)}$ dependence in \eref{effact} to just $N$
for every diagram.  This implies that the matrix diagrams  contributing to the superpotential
are exactly those that survive the large $M$ limit of a bosonic $U(M)$ Matrix model with a potential
\begin{equation}
V(\Phi) = g_2 \Tr(\Phi^2) + g_4 \Tr(\Phi^4) + {\widetilde{g}_2 \over M} \Tr(\Phi^2) \Tr(\Phi^2).
\end{equation}
In Sec.\ 4 we will compute the large $M$ limit of a such a Matrix model and compute the free energy $\CF_0$ in this way.

Below we will compute this effective action (\ref{effact}) to the first few orders.
In Sec.~3 we will show that it is reproduced by an analysis based on
the Seiberg-Witten solution of $\CN = 2$ gauge theories.  In the
single-trace case Dijkgraaf and Vafa argued that the large-$N$ limit of
an associated Matrix model carries out the sum in (\ref{effact}), or
equivalently, that the matrix model free energy provides a generating
function for the perturbative series of matrix  diagrams contributing
to the exact field theory superpotential.    In Sec.\ 4 we will show
that the well known double-trace Matrix models that have large-$N$
limits do sum up the same ``planar pasted diagrams'' that we described
above and give the free energy defined by \eref{free0}.
  However, unlike the single-trace case, the Matrix model will
not reproduce the the combinatorial factors $C(G)$ appearing in
(\ref{effact}). 

\subsection{Perturbative Calculation}

Thus equipped, let us begin our explicit perturbation calculations.
We shall tabulate all
combinatoric data of the pasting diagrams up to third order. Here
$C(G)=\prod_i h_i$ 
and ${\cal F(G)}$ is obtained by counting the contractions of
$\Phi$s. For pure single-trace diagrams the values of ${\cal F(G)}$ have been
computed in Table 1 in \cite{Brezin}, so we can utilize their results.

\subsubsection{First Order}
To first order in coupling constants, all
primitive (diagram (b)) and pasting
diagrams (diagram (a)) are presented in \fref{g2-1}.
Let us illustrate by showing the computations for (a). There is a total
of four index loops and hence $h=4$ for this diagram. Moreover, since it
is composed of the pasting of two primitive diagrams each of which has
$h=2$; thus, we have $C(G) = 2\times 2 = 4$. Finally, $\CF =
\widetilde{g}_2$ because there is only one contraction possible, {\em viz},
$\Tr (\overbracket{\Phi \Phi}) \Tr( \overbracket{\Phi \Phi})$.

\EPSFIGURE[h]{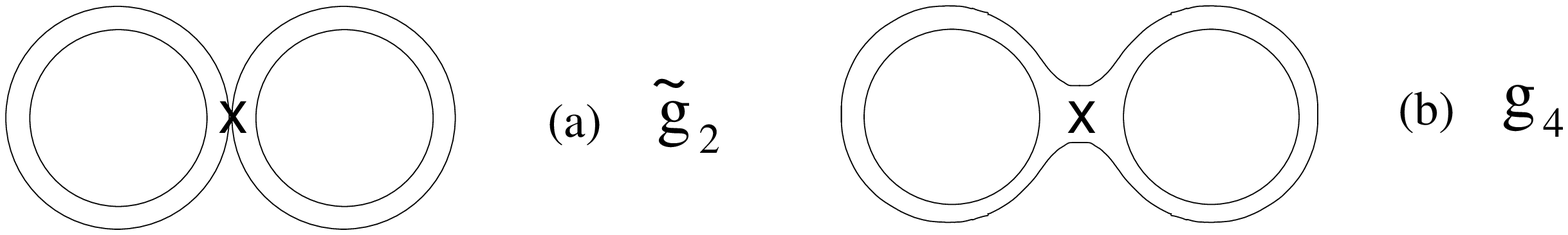,width=13cm}
{All two-loop primitive and pasting diagrams. The vertices have been
marked with a cross.
\label{g2-1}
}
In summary we have:
\beq\label{two}\begin{tabular}{|c|c|c|}\hline
    diagram  &  $(a)$  &  $(b)$
\\ \hline
$h$ &  $4$ & $3$ \\  \hline
$C(G)$ & $4$ &  $3$ \\
\hline ${\cal F}(G)$ & $\widetilde{g}_2$ &  $2g_4$ \\
\hline
\end{tabular}
\eeq

\subsubsection{Second Order}
To second order  in the coupling all primitive ((c)
and (d)) and
pasting diagrams ((a) and (b))
are drawn in \fref{g2-2} and the combinatorics are summarized in table
\eref{three}. Again, let us do an illustrative example. Take diagram
(b), there are five index loops, so $h=5$; more precisely it is composed of
pasting a left primitive diagram with $h=3$ and a right primitive with
$h=2$, so $C(G) = 2 \times 3 = 6$. Now for $\CF(G)$, we need
contractions of the form 
$\Tr (\overbracket{\Phi \Phi} \; \overbracket{\Phi \overbracket{\Phi)
\Tr (\Phi} \Phi})  
\Tr (\overbracket{\Phi \Phi}) $; there are $4 \times 2 \times 2 = 16$
ways of doing so. Furthermore, for this even overall power in the
coupling, we have a minus sign when expanding out the
exponent. Therefore $\CF(G) = -16 \widetilde{g}_2 g_4$ for this diagram.

\EPSFIGURE[h]{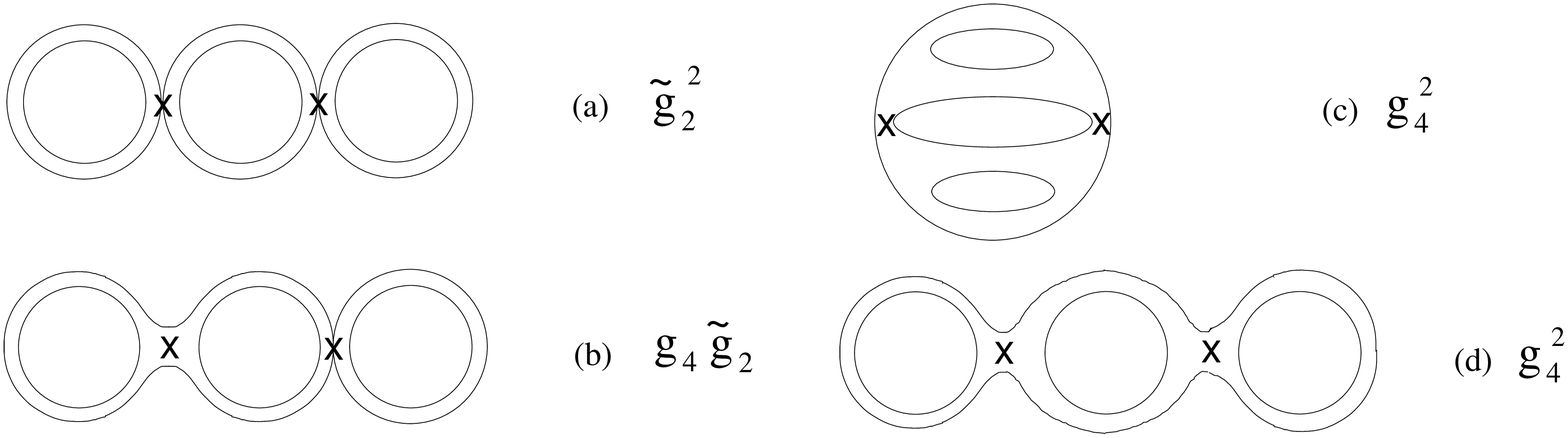,width=15cm}
{All three-loop primitive and pasting diagrams. The vertices have been marked with a
cross.
\label{g2-2}
}

In summary, we have:
\beq\label{three}\begin{tabular}{|c|c|c|c|c|}\hline
    diagram  &  $(a)$  &  $(b)$ & $(c)$ & $(d)$
\\ \hline
$h$ &  $6$ & $5$ & $4$ &$4$  \\  \hline
$C(G)$ & $8$ &  $6$&  $4$ & $4$  \\
\hline ${\cal F}(G)$ & $-4\widetilde{g}_2^2$ &  $-16\widetilde{g}_2g_4$&
$-2g_4^2$ & $-16g_4^2$   \\ 
\hline
\end{tabular}
\eeq

\subsubsection{Third Order}
Finally, the third order diagrams are drawn in \fref{g2-3}. 
The combinatorics are tabulated in \eref{four}.
\EPSFIGURE[t]{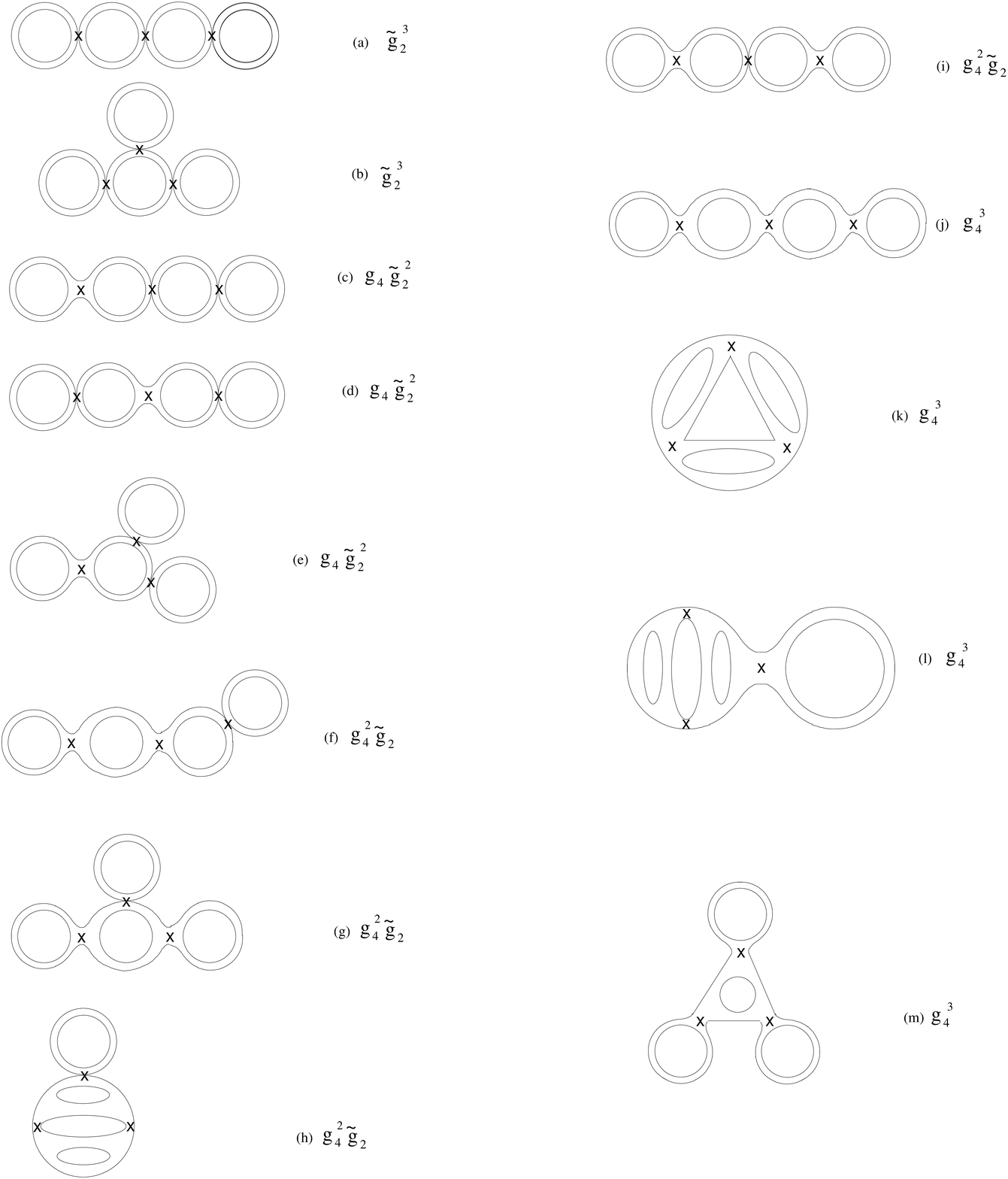,width=15cm}
{All four-loop primitive and pasting diagrams. The
vertices have been marked with a cross.
\label{g2-3}
}
Here the demonstrative example is diagram (b), which is composed of
pasting four diagrams, each with $h=2$, thus $h(G) = 4 \times 2 = 8$
and $C(G) = 2^4 = 16$. For $\CF(G)$, first we have a factor
$\frac{1}{3!}$ from the exponential. Next we have
contractions of the form $\Tr(\overbracket{\Phi \Phi})^3
\Tr(\overbracket{\Phi \overbracket{\Phi) \Tr(\Phi} \; \overbracket{\Phi)
\Tr(\Phi} \Phi})$; there are $2^3 \times 4 \times 2$ ways of doing
this. Thus altogether we have $\CF(G) = \frac{32}{3}\widetilde{g}_2^3$ 
for this diagram.

In summary:
\beq\label{four}\begin{tabular}{|c|c|c|c|c|c|c|c|}\hline
    diagram  &  $(a)$  &  $(b)$ & $(c)$ & $(d)$ & $(e)$ & $(f)$ &
    $(g)$
\\ \hline
$h$ &  $8$ & $8$ & $7$ & $7$ & $7$ & $6$ & $6$  \\
\hline
$C(G)$ &  $16$ & $16$ & $12$ & $12$ & $12$ & $8$ & $8$  \\
\hline
${\cal F}(G)$ &  $16\widetilde{g}_2^3$ & $\frac{32}{3}\widetilde{g}_2^3$ &
    $64\widetilde{g}_2^2g_4$ & 
$32\widetilde{g}_2^2g_4$ & $64\widetilde{g}_2^2g_4$ & $128\widetilde{g}_2g_4^2$ &
    $128\widetilde{g}_2g_4^2$
\\ 
\hline \hline
diagram & $(h)$ & $(i)$ & $(j)$ & $(k)$ & $(l)$ & $(m)$ &
\\ \hline $h$ & $6$ & $6$ & $5$ & $5$ & $5$ & $5$ & \\ \hline
$C(G)$ & $8$ & $9$ & $5$ & $5$ & $5$ & $5$ & \\ \hline ${\cal
F}(G)$ & $32\widetilde{g}_2g_4^2$ & $64\widetilde{g}_2g_4^2$ & $128g_4^3$
& $\frac{32}{3}g_4^3$ & $64g_4^3$ & $\frac{256}{3}g_4^3$ &\\
\hline
\end{tabular}
\eeq

\subsubsection{Obtaining the Effective Action}
Now to the highlight of our calculation.
 From tables (\ref{two}), (\ref{three}), and (\ref{four}) 
we can readily compute the effective
glueball superpotential and free energy. We do so by summing the
factors, with the appropriate powers for $S$, in accordance with
(\ref{effact},\ref{free0}).

We obtain, up to four-loop order,
\begin{eqnarray} \label{free1}
{\cal F}_0
&=& \sum_{G=all~diagrams}{\cal F}(G)S^{h(G)} \nonumber \\
&=&(2g_4+\widetilde{g}_2S)S^3-2(9g_4^2+8g_4\widetilde{g}_2S+2\widetilde{g}_2^2S^2)S^4
\nonumber \\
&+&\frac{16}{3}(54g_4^3+66g_4^2\widetilde{g}_2S+30g_4\widetilde{g}_2^2
S^2+5\widetilde{g}_2^3S^3)S^5+\cdots,
\end{eqnarray}
and subsequently,
\begin{eqnarray} 
\label{glue1}
W_{eff} &=& -NS(\log (S/\Lambda^2)-1)+\sum_{G=all~diagrams}C(G){\cal
F}(G)N^{h(G)-l(G)}S^{l(G)} \nonumber \\
&=&-NS(\log
(S/\Lambda^2)-1)+(6g_4+4\widetilde{g}_2N)NS^2-
(72g_4^2+96g_4\widetilde{g}_2N+32\widetilde{g}_2^2N^2)NS^3 \nonumber \\ 
&& + \frac{20}{3}(6g_4+4\widetilde{g}_2N)^3NS^4+\cdots.
\end{eqnarray}

We will later see how this result may be reproduced from independent
considerations,
{\em i.e.}, the effective action from the factorization of Seiberg-Witten
curve and free energy from the corresponding matrix model.

\subsection{Multiple Traces, Pasted Diagrams and Nonlocality}

Above we found that the only diagrams that contribute to the effective superpotential have zero momentum flowing though the double-trace vertex.  Now observe that the double-trace term in the tree-level action can be written in momentum space as:
\begin{eqnarray}
V &=& \int d^4x \,  \Tr(\Phi^2(x)) \Tr(\Phi^2(x)) \nonumber \\
&=&
\int d^4p_1 \, d^4p_2 \, d^4p_3\, d^4p_4 \,  \Tr(\Phi(p_1)\Phi(p_2)) \Tr(\Phi(p_3)\Phi(p_4)) \, \delta(p_1 + p_2 + p_3 + p_4) .
\end{eqnarray}
Since no momentum flows through the double-trace vertices contributing to the superpotential, the delta function momentum constraint factorizes in our pasted diagrams as
\beq
\delta(p_1 + p_2 + p_3 + p_4) \sim \delta(p_1 + p_2) \delta(p_3 + p_4) .
\eeq
Therefore for the purposes of computing the superpotential we might as well replace the double-trace term in the action by
\begin{eqnarray}
\widetilde{V} &=& 
\int d^4p_1 d^4p_2 d^4p_3 d^4p_4 \Tr(\Phi(p_1)\Phi(p_2)) \Tr(\Phi(p_3)\Phi(p_4)) \, \delta(p_1 + p_2) \ \delta(p_3 + p_4) \nonumber \\
&=&
\int d^4x \, d^4y \, \Tr(\Phi^2(x)) \Tr(\Phi^2(y)) .
\label{nonlocal}
\end{eqnarray}
The Feynman diagrams of this {\it nonlocal} theory include the ones that compute the superpotential in the double-trace, local theory.

This fact suggests that correlation
functions of chiral operators are position independent, as described in \cite{CDSW}.   Along with cluster decomposition, this position independence 
leads to the statement that correlators of operators in the chiral ring of an $\CN = 1$ theory factorize, which we use in the next section to write $\langle \Tr(\Phi^2)^2 \rangle = \langle \Tr(\Phi^2) \rangle^2$.  However it is subtle to establish the precise equivalence between factorization of chiral operators 
and the vanishing that  we demonstrated of all except the pasted diagrams, which leads in turn to the nonlocal action in (\ref{nonlocal}).   We leave this potential connection for exploration in future work.

\section{The Field Theory Analysis}

In this section, we will show that in the confining vacuum the effective
superpotential of the field theory discussed in the previous section is 
\begin{equation} \label{example}
W_{eff}=N\Lambda^2+(6Ng_4+4N^2\widetilde{g}_2)\Lambda^4.
\end{equation}
After integrating in the glueball superfield and expanding the
superpotential in a power series in $S$, (\ref{example})
can be written 
\begin{equation} \label{glue}
W_{eff}=-NS(\log(S/\Lambda^2)-1)+
(6g_4+4\widetilde{g}_2N)NS^2-2(6g_4+4\widetilde{g}_2N)^2NS^3+\frac{20}{3}(6g_4+4\widetilde{g}_2N)^3NS^4+\cdots.
\end{equation}
We shall compare this expression for the low-energy gauge dynamics to the
perturbative field theory computations in Sec.\ \ref{derivation}.  The
two results, of course, are in concert.

We begin by considering an ${\cal N}=1$ $U(N)$ gauge theory with a single
adjoint superfield $\Phi$ deformed from ${\cal N}=2$
by the tree-level superpotential ($n <N$) 
\beq \label{Filed_W}
W_{tree}= \sum_{r=1}^{n+1} g_r u_r + 4\widetilde{g}_2 u_2^2,
\eeq
where 
\beq
u_k := \frac{1}{k}\Tr(\Phi^k). 
\eeq
The tree-level superpotential in (\ref{Filed_W}) is more general than
the one used in (\ref{tree}).  Here, we allow single-trace terms
at arbitrary powers of $\Phi$. We shall
specialize to the previous example at the end of our discussion.

\subsection{The Classical Vacua}
To find the classical vacua, we have to solve the D-term and F-term 
conditions. The D-term is proportional to  $\Tr[\Phi,\bar{\Phi}]^2$ which
is zero if $\Phi$ is diagonal. 
Let the diagonal entries be 
$x_i,~i=1,\ldots, N$. We still need to solve the F-term condition. 
In terms of the $x_i$, the tree-level superpotential is 
\beq
W=\sum_{r=1}^{n+1} {g_r \over r} \sum_{i=1}^N x_i^r +
\widetilde{g}_2 \left(\sum_{i=1}^N x_i^2\right)^2.
\eeq
 From this, the F-flatness condition reads:
\beq
\label{f-flat}
0 = {\partial W\over \partial x_k} =
\sum_{r=1}^{n+1} g_r x_k^{r-1}+4\widetilde{g}_2 x_k \left(\sum_{i=1}^N x_i^2\right), \qquad k=1,...,N.
\eeq
This is certainly different from the case without the double-trace term,
where the F-term equations for different $x_k$s decouple.  Here, the
eigenvalues interact with each other even at the classical level.

To solve the \eref{f-flat}, which may be recast as
\beq\label{eq:mft}
{1\over x_k} 
\sum_{r=1}^{n+1} g_r x_k^{r-1}=-4\widetilde{g}_2  \left(\sum_{i=1}^N x_i^2\right),
\eeq
we take the  RHS of (\ref{eq:mft})
\beq
C :=  4\widetilde{g}_2 \left(\sum\limits_{i=1}^N x_i^2\right) , 
\eeq
as an unknown constant for all $N$ F-terms.
This gives
\beq
\sum_{r=1}^{n+1} g_r x_k^{r-1}+C x_k=0 \qquad \forall\ k.
\eeq
Now the F-terms are decoupled.  We can solve this system just as we solve for
the vacua of a field theory with only single-trace interactions \cite{Cachazo1}
simply by taking $g_2 \mapsto g_2+C$.

As the F-terms are order $n$ polynomials in $x$, we should generically
expect $n$ solutions for each eigenvalue $x_k$.
The eigenvalues are the roots of the polynomial
\beq
0 = \sum_{r=1}^{n+1} g_r x_k^{r-1} \equiv g_{n+1} \prod_{i=1}^n (x-a_i).
\eeq
If $N_i$ of the eigenvalues are located at $a_i$, where $\sum_i N_i = N$,
the unbroken gauge symmetry is 
\beq
U(N) \rightarrow \prod_{i=1}^n U(N_i).
\label{gauge}
\eeq
As $a_i$s are a function of $C$, we need to impose the {\em additional}
consistency condition that 
\begin{equation}  \label{consistent}
4\widetilde{g}_2 \sum_{j=1}^n N_i^2 a_i^2=C.
\end{equation}

To simplify the discussion, we henceforth focus on the special case
where all of the $x_i$s have the same value.
The $SU(N)$ part of the gauge group is unbroken, and will confine in
the infrared. 

\subsection{The Exact Superpotential in the Confining Vacuum} 

We now proceed to find the exact superpotential in this
confining vacuum \cite{SW}. Let us recall the general philosophy of the method
(see {\em e.g.}, \cite{Ferrari:2002jp}, whose notations we adopt, for a
recent discussion).
A generic point in the moduli space of the  $U(N)$  ${\cal N}=2$ theory will be
lifted by the addition of the 
the general superpotential \eref{Filed_W}. The points which are not lifted
are precisely where at least $N-n$ mutually local monopoles become massless.
This can be seen from the following argument. The gauge
group in the ${\cal N}=1$ theory is broken down to $\prod_{i=1}^n U(N_i)$, and the $SU(N_i)$ factors
each confine. We expect  condensation of $N_i -1$ magnetic monopoles 
in each of  these $SU(N_i)$ factors and a total of $N-n$ condensed magnetic monopoles. 
These monopoles condense at the points on the ${\cal N}=2$ moduli space where
 $N-n$ mutually local monopoles become massless. These are  precisely the points which are not lifted by addition of the superpotential. 
These considerations are equivalent to the requirement that the corresponding 
Seiberg-Witten curve has the factorization
\begin{equation}
P_N(x,u)^2-4 \Lambda^{2N}=H_{N-n}(x)^2 F_{2n}(x),
\end{equation}
where $P_N(x,u)$ is an order $N$ polynomial in $x$ with coefficients
determined by the (vevs of) the $u_k$, $\Lambda$ is an ultraviolet cut-off,
and $H$ and $F$ are, respectively, order $N-n$ and $2n$ polynomials in $x$.

The $N-n$ double roots place $N-n$ conditions on the original variables
$u_k$. We can parametrize all the $\langle u_k\rangle $ by $n$ independent
variables $\alpha_j$.  In other words, the $\alpha_j$s then correspond
to massless fields in the low-energy effective theory.  If we know
the exact effective action for these fields, to find the vacua, we simply
minimize $S_{eff}$. Furthermore, substituting $\langle u_k \rangle$ back
into the effective action gives the action for the vacua.

Holomorphy and
regularity of the superpotential as the couplings go to zero requires that
there are no perturbative corrections to the tree-level superpotential. In
addition, we assume that all non-perturbative effects are captured in
the Seiberg-Witten curve analysis discussed above. Then, we need 
to minimize
\beq \label{exact_W}
W_{exact}= \sum_{r=1}^{n+1} g_r \langle u_r\rangle + 4\widetilde{g}_2 
\langle u_2^2 \rangle.
\eeq
In general the factorization problem is hard to solve \cite{Cachazo4},
but for the confining vacuum where all $N-1$ monopoles have condensed,
there is a general solution given by Chebyshev polynomials.\footnote{This
was worked out first by Douglas and Shenker \cite{DS}, but here we use
the results and nomenclature of Ferrari \cite{Ferrari:2002jp}.}
In our case, we have the solution
\newcommand{\Cpq}[2]{{\left( \begin{array}{c} {#1} \\ {#2} \end{array} \right)}}
\begin{eqnarray}
\label{confsol}
&& u_p={N\over p} \sum_{q=0}^{[p/2]} C_{p}^{2q} C_{2q}^q \Lambda^{2q} z^{p-2q}, \\
&& z=\frac{u_1}{N}, ~~~~~~~~~~~~~ C_n^p := \Cpq{n}{p} = \frac{n!}{p!(n-p)!}.
\end{eqnarray}
Notice that in \eref{confsol}, there is one free parameter $z$ which
is the field left upon condensation.
Now we put it into the superpotential
\begin{equation}
W = \sum_{r=1}^{n+1} g_r u_r(z,\Lambda) + 4\widetilde{g}_2  u_2(z,\Lambda)^2,
\end{equation}
solve and back-substitute $z$ from $\partial W/\partial z=0$ to
obtain the effective superpotential $W_{eff}$. Notice that in the above 
result, we have used
\beq
\langle u_2^2 \rangle=\langle u_2 \rangle^2.
\eeq
This is true because $u_2$ is a chiral field, and cluster decomposition
in the field theory lets us factor the correlation functions of operators
in the chiral ring \cite{CDSW}.

Although the above procedure finds $W_{eff}$, it is not the best form to
compare with our previous results because there is no gluino condensate
$S$. To make the comparison, we need to ``integrate in'' \cite{Intri}
the glueball superfield as in \cite{Ferrari:2002jp}.

The integrating in procedure is as follows (here we use the single-trace 
superpotential as an illustrative example of the technique). \\
$\bullet$ We set $\Delta := \Lambda^2$, and use the equation 
\beq
N S=\Delta {\partial W\over \partial \Delta}
=N\sum_{r=2}^{n+1} g_r \sum_{q=1}^{[p/2]} {q\over p} C_{p}^{2q} C_{2q}^q
z^{p-2q} \Delta^q
\eeq
to solve for $\Delta$ in terms of $S$. \\ 
$\bullet$ Next, we find $z$ by solving
\begin{equation}
0={\partial W\over \partial z}=\sum_{r=1}^{n+1} g_r \sum_{q=0}^{[p/2]} 
{p-2q\over p} C_{p}^{2q} C_{2q}^q
z^{p-2q-1} \Delta^q.
\end{equation} \\
$\bullet$ Now the effective action for the glueball superfield $S$ can be written as
\beq
W_{S}(S,g,\Lambda)= -S \log\left({\Delta \over \Lambda^2}\right)^N +
W_{tree}(S, g, \Lambda),
\eeq
which will reproduce the result
\begin{equation}
{\partial \over \partial S}W_S(S,\Lambda^2, g) = -\ln(\Delta/\Lambda^2)^N.
\end{equation}

\subsection{An Explicit Example}
Let us work out the double-trace example that we are interested in solving. The 
superpotential is  
\begin{equation}
\label{g2g4}
W = g_2 u_2 + 4g_4 u_4 + 4\widetilde{g}_2 u_2^2,
\end{equation}
(later, we can set $g_2=m=1$).
Using 
\begin{eqnarray}
u_2&=&{N\over 2}[z^2+2\Lambda^2],\\
u_4&=&{N\over 4}[z^4+12 \Lambda^2 z^2+6 \Lambda^4],
\end{eqnarray}
from \eref{confsol}, we obtain 
\begin{eqnarray}
W & = & N z^2[ {g_2\over 2}+g_4 z^2+12 g_4 \Lambda^2
+4N \widetilde{g}_2  \Lambda^2+N \widetilde{g}_2 z^2] \\
& & + N\Lambda^2[g_2 +6 g_4 \Lambda^2+ 4N\widetilde{g}_2 \Lambda^2]. \nonumber
\end{eqnarray}
 From this we have following equations by setting $\Lambda^2=\Delta$:
\begin{eqnarray}
S & = & \Delta [ g_2+(12 g_4 +4N \widetilde{g}_2)z^2+ (12 g_4 +8N\widetilde{g}_2) \Delta], \\
0 & = & {\partial W \over \partial z}= z[g_2+z^2(4g_4+4N \widetilde{g}_2) +2\Delta (12g_4+4N \widetilde{g}_2)].
\end{eqnarray}
We solve $z=0$:
\beq \label{solve-delta}
\Delta ={-g_2+\sqrt{ g_2^2+4 S(12 g_4+ 8N \widetilde{g}_2)}\over 2 
(12 g_4+ 8N \widetilde{g}_2)}.
\eeq
The effective action with $S$ integrated in is\footnote{Notice that
in both formula \eref{solve-delta} and \eref{W_S}, $g_4$ and
$\widetilde{g}_2$  combine together as $g_4+{2N\over
3}\widetilde{g}_2$. So if we shift $g_4$ to $g_4+{2N\over
3}\widetilde{g}_2$, the single-trace result will reproduce the 
double-trace result, and the effective action for the double-trace
can be naively calculated from the DV prescription by partial 
differentiation of the glueball field $S$.} 
\beq  \label{W_S}  
W_{eff}  =  S \log \left({\Lambda^{2} \over \Delta}\right)^N + 
N \Delta[g_2 + \Delta  (6 g_4+ 4N \widetilde{g}_2)],
\eeq
which will be the one used in comparison to our previous results.
After minimizing this action, we find 
\beq
\label{unscaled}
W_{z=0}=N\Lambda^2[g_2 +6 g_4 \Lambda^2+ 4N\widetilde{g}_2
\Lambda^2],
\eeq
which is the promised result of \eref{example}.
Setting $g_2=1$, expanding \eref{solve-delta} in powers of
$S$, and substituting into \eref{W_S}, we get the second formula 
\eref{glue} from the beginning of this section:
\begin{equation} 
W_{eff}=-NS(\log
(S/\Lambda^2)-1)+(6g_4+4\widetilde{g}_2N)NS^2-2(6g_4+4\widetilde{g}_2N)^2NS^3+
\frac{20}{3}(6g_4+4\widetilde{g}_2N)^3NS^4+\cdots. 
\end{equation}
Crucial in matching the result of this calculation with the
perturbative analysis in the previous section is the assumption of
factorization $\langle u_2^2 \rangle=\langle u_2 \rangle^2$. This is
equivalent to the vanishing the pinching diagrams in the perturbative
analysis.

\section{The Matrix Model}

In Sec.\ 2 we demonstrated that the explicit field theory computation
of the effective superpotential localizes to a certain sum of matrix
diagrams.   All of these diagrams are constructed by pasting planar
single-trace diagrams together with double-trace vertices in such a
way that no additional momentum loops are created.  By examining the
scaling of these diagrams with $N$ in (\ref{effact}) we also observed
that these are precisely the diagrams that would survive the large-$M$
limit of a $U(M)$ bosonic Matrix model with a potential 
\begin{equation}
V(\Phi) = g_2 \, \Tr(\Phi^2) + g_4 \Tr(\Phi^4) + {\widetilde{g}_2 \over M}
\Tr(\Phi^2)\Tr(\Phi^2).
\label{doubletrace} 
\end{equation}
The extra factor of $1/M$ multiplying $\widetilde{g}_2$, in comparison
with the field theory tree-level superpotential (\ref{doubtrace}) is
necessary for a well-defined 't Hooft large-$M$ limit.   This is
because each trace, being a sum of eigenvalues,  
will give a term proportional to $M$.  So to prevent the double-trace
term from completely dominating the large-$M$ limit we must divide by
an extra factor of $M$.  Fortunately, this a  
well known model and was solved more than a decade ago  \cite{Das,
Aki}.  Below we will review this solution, compare its results to our
field theory calculations, and then generalize to other multi-trace
deformations.

\subsection{The Mean-Field Method}
\label{DDSW} 

The basic observation, following \cite{Das}, that allows us to solve
the double-trace matrix model (\ref{doubletrace}), is that in the
large-$M$ limit the effects on a given matrix eigenvalue of all the
other eigenvalues can be treated in a mean field approximation.
Accordingly we compute the matrix model free energy $\CF$ as 
\begin{eqnarray}
\label{5.1} \exp(-M^2 {\cal F}) &=& 
\int d^{M^2}(\Phi)\exp\{-M(\frac{1}{2}\Tr(\Phi^2)+
g_4\Tr(\Phi^4)+\widetilde{g}_2\frac{(\Tr(\Phi^2))^2}{M})\}.\\
\nonumber 
&=& 
\int \prod_i d\lambda_i \exp\{
M(-\frac12 \sum_i \lambda_i^2 - \frac{g_4}{M} \sum_i \lambda_i^4 -
\frac{\widetilde{g}_2}{M^2}(\sum_i \lambda^i)^2) + \sum_{i \ne j} \log
|\lambda_i - \lambda_j| \}
\end{eqnarray}
Here $\lambda$ are the $M$ eigenvalues of $\Phi$ and
$\cal F$ is the free energy, which can be evaluated by saddle
point approximation at the planar limit. The $\log$ term comes from
the standard Vandermonde determinant.
This matrix model is
Hermitian with rank $M$ in the notation of \cite{DV4} (of
course as mentioned earlier, we should really consider $GL(M,\IC)$
matrices though the techniques hold equally).   We have introduced an
extra factor of $M$ in the exponent on the right hand side of
(\ref{5.1}) by rescaling the fields and couplings in
(\ref{doubletrace}) in accordance with the conventions of \cite{Das}. 

The density of eigenvalues
\beq
\rho(\lambda):=\frac{1}{M}\sum_{i=1}^{M}\delta(\lambda-\lambda_i)
\eeq
becomes continuous in an interval $(-2a,2a)$ when $M$ goes to
infinity in the planar limit for some $a \in \IR^+$.
Here the interval is symmetric
around zero since our model is an even function. The normalization
condition for eigenvalue density is
\begin{equation} \label{5.2}
\int_{-2a}^{2a}d\lambda \rho(\lambda)=1 \ .
\end {equation}
We can rewrite (\ref{5.1}) in terms of the eigenvalue density in the
continuum limit as
\begin{eqnarray} \label{5.11}
\exp(-M^2 {\cal F})&=&\int \prod _{i=1}^{M} d\lambda_i
\exp\{-M^2(\int_{-2a}^{2a} d\lambda \rho(\lambda)(
\frac{1}{2}\lambda^2+g_4\lambda^4) \\ \nonumber
&+&\widetilde{g}_2(\int_{-2a}^{2a} d\lambda
\rho(\lambda)\lambda^2)^2-\int_{-2a}^{2a}\int_{-2a}^{2a} d\lambda
d\mu \rho(\lambda)\rho(\mu)\ln|\lambda-\mu|)\} \ .
\end {eqnarray}
Then the saddle point equation is
\begin{equation} \label{5.3}
\frac{1}{2}\lambda+2g_4\lambda^3+2\widetilde{g}_2c\lambda=P
\int_{-2a}^{2a}d\mu\frac{\rho(\mu)}{\lambda-\mu},
\end{equation}
where $c$ is the second moment
\begin{equation} \label{5.4}
c := \int_{-2a}^{2a}d\lambda \rho(\lambda)\lambda^2
\end{equation}
and $P$ means principal value integration.

The effect of the double-trace is to modify the coefficient of
$\lambda$ in the saddle point equation. We can determine the
number $c$ self-consistently by (\ref{5.4}). The solution of
$\rho(\lambda)$ to (\ref{5.3}) can be obtained by standard matrix model
techniques by introducing a resolvent. The answer
is
\begin{equation} \label{5.5}
\rho(\lambda)=\frac{1}{\pi}\left(\frac{1}{2}+2\widetilde{g}_2c+
4g_4a^2+2g_4\lambda^2\right)\sqrt{4a^2-\lambda}.
\end {equation}
Plugging the solution into (\ref{5.2}) and (\ref{5.4}) we
obtain two equations that determine the parameters $a$ and $c$:
\begin{equation} \label{5.6}
a^2(1+4\widetilde{g}_2c)=1-12g_4a^4,
\end{equation}
\begin{equation} \label{5.7}
16g_4\widetilde{g}_2a^8+(12g_4+4\widetilde{g}_2)a^4+a^2-1=0.
\end{equation}

Substituting these expressions into (\ref{5.11}) gives us the
free energy in the planar limit $M \rightarrow \infty$ as:
\begin{eqnarray}
{\cal F}&=&\int_{-2a}^{2a} d\lambda \rho(\lambda)(
\frac{1}{2}\lambda^2+g_4\lambda^4)
+\widetilde{g}_2c^2-\int\int_{-2a}^{2a} d\lambda d\mu
\rho(\lambda)\rho(\mu)\ln|\lambda-\mu|.\nonumber \\
\end{eqnarray}
One obtains
\begin{equation} \label{freeenergy}
{\cal F}(g_4,\widetilde{g}_2)-{\cal
F}(0,0)=\frac{1}{4}(a^2-1)+(6g_4a^4+a^2-2)g_4a^4-\frac{1}{2}\log
(a^2) \ .
\end{equation}
Equation (\ref{freeenergy}) together with (\ref{5.7}) give the
planar free energy. We can also expand the free energy in powers of
the couplings, by using 
(\ref{5.7}) to solve for $a^2$ perturbatively
\begin{eqnarray}
a^2&=&1-(12g_4+4\widetilde{g}_2)+(288g_4^2+176g_4\widetilde{g}_2+32\widetilde{g}^2_2)\\
\nonumber
&&- (8640g_4^3+7488g_4^2\widetilde{g}_2+2496g_4\widetilde{g}_2^2
+320\widetilde{g}_2^3)+\cdots.
\end{eqnarray}
Plugging this back into (\ref{freeenergy}) we find the free energy as
a perturbative series 
\begin{equation} \label{free}
\CF_0 = {\cal F}(g_4,\widetilde{g}_2)-{\cal
F}(0,0)=2g_4+\widetilde{g}_2-2(9g_4^2+8g_4\widetilde{g}_2+2\widetilde{g}_2^2)
+\frac{16}{3}(54g_4^3+66g_4^2\widetilde{g}_2+30g_4\widetilde{g}_2^2+
5\widetilde{g}_2^3)+\cdots.
\end{equation}

Comparing with \eref{free1} we see that $\CF$ reproduces the explicit
computation of the generating function of  ``planar pasted" field
theory diagrams in Sec.\ 2.5.  In matching the two we have to restore
the proper powers of  
the glueball $S$ into \eref{free}. 
First recall that to keep the relevant diagrams in the matrix
model, we have inserted ${1\over M}$ to the double-trace term
in \eref{doubletrace}. Therefore to
compare with  \eref{free1}, we need to rescale $\widetilde{g}_2$ in 
\eref{free} to $\widetilde{g}_2 M\equiv\widetilde{g}_2 S$ where we have
effectively identified the glueball $S$ in the field theory with $M$
in the matrix model. 
In addition, we should re-insert powers of $M$ into \eref{free} by
loop counting. 
The first two terms in \eref{free} have three index loops so we need
to multiply them by $M^3 \equiv S^3$. 
The third term has four index loops and fourth term,
five and hence we respectively need factors of $S^4$ and $S^5$.
With these factors correctly placed into \eref{free}, we
recover \eref{free1} completely.

This verifies our claim that the diagrams surviving the 
large-$M$ limit of the matrix model (\ref{doubletrace}) are precisely
the graphs that contribute to  
effective action of the field theory with the tree-level
superpotential (\ref{doubtrace}).   Nevertheless, as we have already
discussed in Sec.\ 2.4 we cannot compute the effective superpotential
of the field theory $W_{eff}(S)$ by taking a derivative $\partial
\CF_0 /\partial S$, because the combinatorial  
factors will not agree.     In this way the double-trace theories
differ in a significant way from the single-trace models discussed by
Dijkgraaf and Vafa \cite{DV3}.  We can pose the challenge of finding the field
theories who effective superpotential is computed by the Matrix model
(\ref{freeenergy}).

%
\subsection{Generalized Multi-Trace Deformations}\label{gendef}

In fact the mean field techniques of the previous subsection can be generalized to solve the general multi-trace model.   Below we illustrate this by solving the general quartic Matrix model; as discussed above, it is an interesting challenge to find a find a field theory whose effective superpotential these models compute.

Specifically, let us consider the Lagrangian
\begin{equation}
\label{genmulti}
{\cal L} = g_2\Tr(\Phi)^2 + \mu (\Tr(\Phi))^2 + \nu_1 (\Tr\Phi)^2(\Tr(\Phi^2))
+ \nu_2 (\Tr\Phi)^4 + 2 \nu_3 (\Tr\Phi)(\Tr(\Phi^3)),
\end{equation}
which exhausts all quartic interactions.

The one-matrix model partition function
\begin{equation}
Z_M = \int [D\lambda]\, \exp\left\{-M^2\left({\cal L} -
\int_0^1 dx\, \int_0^1 dy\, \log|\lambda(x)-\lambda(y)|\right)\right\},
\end{equation}
gives the saddle point equation
\begin{equation}
\label{newsaddle}
4 g_4 \lambda^3 + 6 \nu_3 c_1 \lambda^2 +
2(g_2 + 2 \widetilde{g}_2 c_2 + \nu_1 c_1^2) \lambda +
2(\mu c_1 + \nu_1 c_1 c_2 + 2 \nu_2 c_1^3 + \nu_3 c_3) =
2 P \int_{-2a}^{2b} d\tau\, \frac{u(\tau)}{\lambda-\tau},
\end{equation}
where the moments $c_k$ are defined as
\begin{eqnarray}
\label{moments}
c_k &=& \int_{-2a}^{2b} d\tau\, u(\tau) \tau^k, \nonumber \\
c_0 &=& \int_{-2a}^{2b} d\tau\, u(\tau) = 1.
\end{eqnarray}

Note that we have introduced the separate upper and lower cut
parameters $a$ and $b$ as opposed to the standard symmetric treatment
because $u(\lambda)$ is not of explicit 
parity (such asymmetric examples have also been considered in
\cite{Brezin}). 
When $a=b$ one can  recast \eref{newsaddle} into
a Fredholm integral equation of the
first kind and Cauchy type, which affords a general solution as
follows \cite{inteq}
\beq
P \frac1{\pi} \int_{-a}^a \frac{u(t)dt}{t-x} = -v(x) \Rightarrow
u(x) = \frac{1}{\pi} \int_{-a}^a
\left( \sqrt{\frac{a^2 - t^2}{a^2 -x^2}} \right)
\frac{v(t) dt}{t-x} + \frac{C}{\sqrt{a^2-x^2}}
\eeq
for some constant $C$.   When $a \neq b$ we can use the ansatz:
\begin{equation}
u(\lambda) =
\frac{1}{\pi}(A \lambda^2 + B \lambda + C)
\sqrt{(2a+\lambda)(2b-\lambda)}
\end{equation}
with the constants matching the coefficients in the LHS of
\eref{newsaddle} as
\begin{eqnarray}
\label{abcd}
2 A &=& 2g_4, \\ \nonumber
2 a A - 2 A b + 2 B &=& 3 \nu_3 c_1, \\ \nonumber
-a^2\,A  - 2\,a\,A\,b - A\,b^2 + 2\,a\,B -
  2\,b\,B + 2\,C
&=& g_2 + 2\widetilde{g}_2 c_2 + \nu_1 c_1^2, \\ \nonumber
a^3\,A + a^2\,A\,b - a\,A\,b^2 - A\,b^3 - a^2\,B - 2\,a\,b\,B - b^2\,B
 + 2\,a\,C - 2\,b\,C &=&
\mu c_1 + \nu_1 c_1 c_2 + 2 \nu_2 c_1^3 + \nu_3 c_3.
\end{eqnarray}
We see a well-behaved $u(\lambda)$ which is zero at the
end-points and vanishes outside the support $(-2a,2b)$.

We now need to check the consistency of our mean-field method.
This simply means the following. Considering the definition of $c_i$
in \eref{moments}, the definitions \eref{abcd}
actually constitute a system of equations for
$A,B,C,a,b$ because each $c_i$ on the RHS, through
\eref{moments}, depend on $A,B,C,a,b$. To \eref{abcd} we
must append one more normalization condition, that
$c_0 = \int_{-2a}^{2b} u(\lambda) d \lambda = 1$.
Therefore we have five equations in five variables which will fix our
parameters in terms of the seven couplings. Our mean-field method is
therefore self-consistent.   It would be interesting to find a role
for such exactly solvable models in the physics of four-dimensional
field theories.

\section{Linearizing Traces: How To Identify the Glueball?}

In previous sections we showed that the field theory computation of the effective superpotential of a double-trace theory as a function of the glueball $S$ localized to summing Matrix diagrams.   In the end, the only the ``pasted'' diagrams that contributed, namely certain tree-like graphs obtained by pasting together planar single-trace graphs with double-trace vertices in such a way that no momentum flows through the latter.
After verifying the result via the $\CN =2$ Seiberg-Witten solution, we demonstrated that the sum of these diagrams given as a series in (\ref{effact}) is not computed by the large $M$ limit of a $U(M)$ Matrix as one would have naturally hoped.  Nevertheless, we may wonder if there is
some Matrix model that sums the series of pasted diagrams.   In this section we take up the challenge of finding such a Matrix model.

Since the authors \cite{DGLVZ} and \cite{CDSW} have proven that the superpotential of a single-trace gauge theory can be computed from an associated Matrix model, we seek to construct our double-trace theory from another single-trace model.
Recall that we are considering the tree-level superpotential
\begin{equation}
W_{tree}=\frac{1}{2}g_2\Tr(\Phi^2)+g_4\Tr(\Phi^4)+\widetilde{g}_2
(\Tr(\Phi^2))^2.
\label{doubtraceagain}
\end{equation}
Now consider another theory with an additional gauge singlet field $A$ 
\begin{equation} \label{m1.2}
W_{tree}=\frac{1}{2}(g_2+4\widetilde{g}_2A)\Tr(\Phi^2)+g_4\Tr(\Phi^4)-
\widetilde{g}_2A^2.
\label{singtraceA}
\end{equation}
It is easy to see that integrating out $A$ in \eref{singtraceA}, which amounts to 
solving $\diff{W_{tree}}{A} = 0$ and back-substituting, produces the double-trace theory (\ref{doubtraceagain}).

The advantage of
\eref{m1.2} is that it consists purely of single-trace operators. 
The first two terms will generate an effective potential ${W}_{\rm single}(A,S)$, as a function
of $A$ and the glueball superfield $S$ (the subscript ``single'' refers to the fact that this is
the superpotential for the model without the double-trace term and with an $A$ dependent
mass term). Then
\begin{equation}
W_{\rm eff}(A,S) ={{W}}_{\rm single}(A,S)-\widetilde{g}_2 A^2.
\label{weffas}
\end{equation}
The exact superpotential for the glueball superfield $S$ for the double-trace theory then follows by integrating $A$ out, {\em i.e.} solving ${\partial {{W_{\rm single}}} \over \partial A}-2\widetilde{g}_2 A =0$ for $A$ and substituting in (\ref{weffas}).   
Since single-trace theories are directly related to Matrix model we might hope to use this construction with an added auxiliary field $A$ to find an auxiliary Matrix model that sums the pasted diagrams of the double-trace theory.

\subsection{Field Theory Computation of $\mathbf{W_{\rm single}(A,S)}$ and Pasted Matrix Diagrams}

In this section\footnote{We thank Cumrun Vafa and Ken Intriligator for communications concerning the material in this section.} we will discuss how the superpotential for the double-trace theory can be computed in field theory from the linearized model \eref{singtraceA}.   First, observe that the superpotential for an adjoint theory with an additional gauge singlet \eref{m1.2} localizes to summing matrix integrals,
since  the arguments of \cite{DGLVZ} that are reviewed in Sec.~\ref{derivation} go through essentially unchanged.   To compute
the effective potential as a function of both $A$ and the glueball $S$, we need to sum superspace
Feynman diagrams with insertions of both $A$ and $\CW_{\alpha}$, with both of these treated as background
fields.
Since we are only interested in contributions to the superpotential, we can restrict ourselves to constant background $A$.   Then it is easy to verify that the entire analysis in Sec.~\ref{derivation} goes through for the theory \eref{m1.2}, with the double-trace coupling $\widetilde{g}_2$ set to zero and  a shift in the mass of the field $\Phi$, 
{\it viz} $g_2 \rightarrow g_2 + 4\widetilde{g}_2$.  In particular, the computation of the effective superpotential 
${W}_{\rm single}(A,S)$ localizes to summing matrix diagrams and there is some free energy $\CF_{{\rm single}}$ in terms of which $W_{{\rm single}} = N \partial{\cal F_{\rm single}}(S,A)/\partial S$.

Let us verify that this procedure will yield the correct double-trace result when we integrate $A$ out.
Making the $g_2 \rightarrow g_2 + 4\widetilde{g}_2$ with $\widetilde{g}_2 = 0$
in the known single-trace 
result \eref{W_S} and \eref{solve-delta}, we find the
effective superpotential
\begin{equation} 
\label{m1.3}
{W}_{\rm single}(A,S)=NS\log(\frac{\Lambda^2}{\Delta})+
N\Delta((g_2+4\widetilde{g}_2A)+6g_4\Delta),
\end{equation}
where $\Delta$ is determined by the quadratic equation
\begin{equation} \label{m1.4}
12g_4\Delta^2+(g_2+4\widetilde{g}_2A)\Delta=S.
\end{equation}
Now we can integrate $A$  out  and obtain the superpotential for the glueball superfield $S$. 
We solve $\partial W_{\rm eff}/\partial A=\partial {W_{\rm single}}/\partial A-2{\widetilde{g}}_2 A=0$ for $A$.
This is a
simple calculation from \eref{m1.3}, \eref{m1.4}:
\begin{eqnarray}
\frac{\partial}{\partial A}W_{\rm single}(\Delta(S,A),S,A)
&=&\frac{\partial W_{\rm  single}}{\partial A}+
   \frac{\partial \Delta}{\partial A}\frac{\partial W_{\rm single}}{\partial \Delta} \\ \nonumber
&=&(g_2+4\widetilde{g}_2A+12g_4\Delta-\frac{S}{\Delta})N\frac{\partial
\Delta}{\partial A}+4\widetilde{g}_2N\Delta \\ \nonumber 
&=& 4\widetilde{g}_22N\Delta,
\end{eqnarray} 
where in last step we have used \eref{m1.4}.
So the solution to $\partial W_{\rm single}/\partial A-2{\widetilde{g}}_2 A=0$ is 
\begin{equation} 
A=2N\Delta.
\end{equation}
Plugging $A=2N\Delta$ into \eref{m1.3}, \eref{m1.4} and \eref{weffas}, we find the
effective glueball superpotential to be
\begin{equation} \label{m1.9}
W_{\rm  eff}=NS\log(\frac{\Lambda^2}{\Delta})+N\Delta(g_2+4\widetilde{g}_2N\Delta+6g_4\Delta),
\end{equation}
with $\Delta$ determined by the quadratic equation
\begin{equation} \label{m1.10}
(12g_4+8\widetilde{g}_2 N)\Delta^2+g_2\Delta=S.
\end{equation}
Equations (\ref{m1.9}) and (\ref{m1.10}) are of course the double-trace
effective glueball superpotential we computed previously  in  \eref{solve-delta} and  \eref{W_S}.

Why does this procedure reproduce precisely the sum of pasted diagrams that contribute to the double-trace superpotential in \eref{effact}? From the point of view of perturbation theory, we are doing the
following. If we treat $A$ as a constant, we should simply sum
the planar diagrams in the theory with a quartic superpotential, and
after doing so, we obtain the superpotential $\int d^4x d^2\theta\ W$
with
\begin{equation}
W = W_{{\rm connected\,planar}}(S,g_2+ 4 \widetilde{g}_2 A,g_4) - \widetilde{g}_2
A^2.
\end{equation}
Next, we should integrate out $A$. To do so, we write $A=A_0 + \widetilde{A}$,
where $A_0$ solves $\partial W/\partial A=0$. We see that $W$ becomes
\begin{equation}
W = W_{{\rm connected\,planar}}(S,g_2 + 4 \widetilde{g}_2 A_0,g_4) -
\widetilde{g}_2 A_0^2 + c_2 \widetilde{A}^2 + c_3 \widetilde{A}^3 + \ldots.
\end{equation}
What is the meaning of integrating over $\widetilde{A}$? From the diagrammatic
point of view, $\widetilde{A}$ is the field that allows momentum to flow
through
the $\widetilde{g}_2$ vertices. All diagrams where such momentum flow is
prohibited are
taken into account by the background value $A_0$. Thus, picking $A_0$
takes
the pasting process into account, whereas the further integrals over
$\widetilde{A}$ should correspond to pinching diagrams. We already know that
these
latter diagrams should vanish
from our diagrammatic analysis, and therefore we should
simply drop all terms involving $\widetilde{A}$. The final answer for $W$ is
thus
\begin{equation} \label{finaa}
W = W_{{\rm connected\,planar}}(S,g_2 + 4 \widetilde{g}_2 A_0,g_4) -
\widetilde{g}_2 A_0^2.
\end{equation}
One can also see directly that integrating out $\widetilde{A}$ gives no
contribution
to the superpotential. In the diagrams that one can write down, there will
be
many loops of $\widetilde{A}$, but there are no vertices that can absorb any
fermionic
momentum, and therefore these diagrams do not yield any contribution to
the
superpotential.

It is an interesting exercise to verify explicitly that (\ref{finaa}) is a
generating diagram for pasted diagrams (which are all tree graphs), made
out of building blocks corresponding to $W_{{\rm connected\,planar}}$.

\subsection{Matrix model perspective}
Above we argued that the methods of \cite{DGLVZ} show that above linearization of the double-trace deformation
via introduction of an auxiliary singlet field $A$
leads to a theory whose superpotential  be computed by a matrix model.   

First observe that  the  double-trace matrix model partition function can be linearized in traces by the introduction of an auxiliary parameter $A$, over which we integrate:
\begin{eqnarray}
Z = \exp(-M^2 \CF^{\rm double}_0) &=& \int d^{M^2}(\Phi)\exp\{-M(\frac{1}{2}g_2 \Tr(\Phi^2)+
g_4\Tr(\Phi^4)+\widetilde{g}_2\frac{(\Tr(\Phi^2))^2}{M})\} \label{singtracemat1} \\
& = & \int dA ~d^{M^2}(\Phi)\exp\{-M(\frac{1}{2}(g_2+4\widetilde{g}_2 A)\Tr(\Phi^2)+
g_4\Tr(\Phi^4)-M{\widetilde{g}}_2A^2)\}. \nonumber
\end{eqnarray}
This is is Matrix model analog of the statement that the double-trace field theory can be generated by integrating out a gauge singlet.
In terms of the free energy of the single-trace matrix model, this can be written as
\begin{equation}
\exp(-M^2 \CF^{\rm double}_0)=\int dA \exp(-M^2 \CF^{\rm single}_0 + M^2{\widetilde{g}_2}A^2).
\label{doubtracefree}
\end{equation}
Hence to obtain the free energy of the double-trace matrix model, we need to solve
the equation ${\partial \CF^{\rm single}_0(A,S) \over \partial A}-2{\widetilde{g}_2}A=0$ for $A$ and 
substitute in $\CF^{\rm double}_0=\CF^{\rm single}_0(A,S)-\widetilde{g}_2 A^2$ where we have
used the identification from \cite{DV3} that $S \sim M$.
The resulting expression for the double-trace matrix model is, of course, the same as that obtained by mean field methods in section (\ref{DDSW}). However, as emphasized in the previous sections, the derivative of this free energy with respect to $S$ does {\it not} yield the correct superpotential for the field theory.

Let us now contrast this with a {\it different} matrix model construction suggested by the field theory analysis in the previous subsection.  Consider the Matrix partition function
\begin{equation}
\widetilde{Z} = \exp(-M^2 \CF^{\rm single}_0) =  \int ~d^{M^2}(\Phi)\exp\{-M(\frac{1}{2}(g_2+4\widetilde{g}_2 A)\Tr(\Phi^2)+
g_4\Tr(\Phi^4)\},
\end{equation}
where $A$ is now treated as a fixed parameter of the Matrix model in analogy with the constant $A$ appearing in the field theory superpotential.    As we explained above, the arguments of \cite{DGLVZ} applied to the linearized model \eref{m1.2} show that in terms of ${\CF^{\rm single}_0}(A,S)$ (with $S \sim 
M$ as in \cite{DV3})
\begin{equation}
W_{\rm eff}(A,S)= \left. - N S (\log(S/\Lambda^2)-1) + 
N {\partial \CF^{\rm single}_0 \over \partial S}\right|_{{\rm constant A}}-\widetilde{g}_2 A^2,
\end{equation}
where $W_{{\rm eff}}$ is the field theory superpotential in \eref{weffas}.   Note that we have {\it not} 
integrated out $A$ at this stage.  Since the single-trace matrix model has reproduced $W_{{\rm eff}}(A,S)$, it is manifest that 
integrating out $A$ will correctly produce the double-trace superpotential as a function of the glueball, just as it did in the field theory.

\paragraph{Identifying the glueball: }
Something remarkable appears to have happened here.  If we start with the single-trace Matrix model \eref{singtracemat1} and integrate out the singlet $A$, we find the free energy of the double-trace model as indicated in \eref{doubtracefree}, and $\partial\CF^{{\rm double}}_0/\partial S$ does not reproduce the field theory superpotential.   
However, if we first differentiate with respect to $S$ and {\it then} integrate out $A$ we reproduce the field theory result.  Of course, the direct field theory computation described above indicates to us that this is the right order in which to do things.   But notice that the difference between the two procedures essentially amounts to correctly identifying the glueball.   Apparently to identify $S$ in the Matrix model we must
linearize the theory and then use the prescription given in \cite{DV3}.   Of course, differentiating the single-trace free energy with respect to $S$ at constant $A$ translates in the double-trace theory 
into some complicated operation which would in effect identify the field theory glueball in that Matrix model.  But it is challenging to identify what this operation is.

\subsection{General Multi-trace Operators}

Finally, we show that the procedure of introducing auxiliary parameters to linearize traces in a double-trace theory can be extended to a general multi-trace model.   Consider the term
\beq
\label{case2W}
 \Tr (\Phi^{m_1}) \Tr (\Phi^{m_2}). 
\eeq
in the superpotential. 
We can rewrite this in terms of single trace terms by introducing four gauge singlet fields $A_i$, $i=1\cdots 4$ as
\beq
\label{case2}
W_2 = 3 \Bigl(A_1^2+A_2^2+ A_1 A_2 + A_1 \Tr \Phi^{m_1}+ A_2 \Tr (\Phi^{m_2}+{2 \over \sqrt{3}} A_3 \Tr \Phi^{m_1}-A_3^2+{2 \over \sqrt{3}} A_4 \Tr \Phi^{m_2}-A_4^2 \Bigr).
\eeq
Integrating out $A_i$  by setting 
$\partial W_2/\partial A_i= 0$, solving for $A_i$ and substituting in \eref{case2} yields the double trace superpotential \eref{case2W}.

To generate a term of the form
\beq
\label{case3W}
\Tr (\Phi^{m_1}) \Tr (\Phi^{m_2}) \Tr (\Phi^{m_3}) \ .
\eeq 
we iterate the above procedure twice, {\em i.e.} we introduce additional gauge singlet fields $B_i$, $i=1 \cdots 4$, and consider the theory with a superpotential
\beq
W_3=3 \Bigl(B_1^2+B_2^2+ B_1 B_2 + B_1 \Tr \Phi^{m_3}+ B_2 W_2+{2 \over \sqrt{3}} B_3 \Tr \Phi^{m_3}-B_3^2+{2 \over \sqrt{3}} B_4 W_2 -B_4^2 \Bigr).
\eeq
where $W_2$ is defined in \eref{case2}. 
Integrating out $A_i$  and $B_i$ for  $i=1\cdots 4$  yields the term \eref{case3W}. Generalization to terms with more traces is obvious.

%
\section{Conclusion}

We have studied an $\CN = 1$ $U(N)$ gauge theory with adjoint chiral
matter and a double-trace tree-level superpotential.   We found by
direct computation that the computation of the effective
superpotential as a function of the glueball superfield localizes to
summing a set of matrix integrals.  The associated set of Matrix
diagrams have the structure of tree diagrams in which double-trace
vertices are strung together by ``propagators'' and ``external'' legs
are that themselves connected single-trace diagrams.   We showed that
the Seiberg-Witten solution to $\CN = 2$ field theories computes an
effective superpotential for the double-trace theory that matches our
direct analysis.   The use  
of factorization in our Seiberg-Witten analysis, namely that $\langle \Tr(\Phi^2)^2 \rangle = \langle \Tr(\Phi^2) \rangle^2$, was equivalent in our perturbative computations to the vanishing of any diagrams  where extra momentum loops were introduced by the double-trace vertices.
Next, we showed that the large-$M$ limit of the standard double-trace
$U(M)$ Matrix model does sum up the same set of matrix diagrams, but
the combinatorial factors are different from those appearing in the
field theory.  In particular, the field theory superpotential is not
computed by a derivative of the matrix model free energy as in
\cite{DV3}.   Put another way, a simple manipulation of the free
energy of the standard double-trace matrix model does not give a
generating function for  the field theory superpotential. Finally we demonstrated how a multi-trace
field theory can be linearized in traces by the introduction of auxiliary singlet fields.  We showed the associated Matrix model, which is linear in traces also, computes the field theory superpotential as a function of both the glueball and the new singlets.   The basic subtlety, then, lies in the correct identification of the field theory glueball as a variable in a Matrix model.

Our results raise several challenges: 
\begin{enumerate}
\item We found that while a multi-trace Matrix model did not directly compute the superpotential of a multi-trace field theory, we could sum the necessary diagrams by introducing auxiliary singlet fields in an associated single-trace model.  The basic subtlety involves correct identification of the field theory glueball in the Matrix model.   How is this done in general, and what is the underlying principle driving the identification?
\item Does the large-$N$ limit of the standard double-trace Matrix
model compute the superpotential for some $\CN = 1$ field theory? 
\item We expect that all our results generalize easily to multi-trace
theories --- it would be nice to check this. 
\item We have worked in the vacuum with an unbroken gauge group.   It
would be good to generalize our arguments to the other vacua with
partially broken gauge symmetry. 
\item In Sec.\ 2.6 we pointed out that there is an intriguing connection between the contributions made by multi-trace vertices to the superpotential of a local theory and certain Feynman diagrams of an associated {\it nonlocal} theory.   It would be very interesting to flesh out this connection.
\item In an $U(N)$ theory with adjoint $\Phi$, the operator
$\Tr(\Phi^K)$ with $K>N$ decomposes into a sum of multi-trace
operators.   This decomposition can receive quantum corrections as
discussed in \cite{CDSW}.  How do our arguments generalize to this
case?   
\item We can also add baryon-like operators like $\det(\Phi)$ to
the superpotential (for theories with fundamental matter in the
context of matrix models, baryons were studied by 
\cite{Argurio:2002hk,Bena:2002ua}).  
Such operators also decompose into sums of
products of traces, and are very interesting because, even without
fundamental matter, they can give rise to an {\it open} string sector
in Yang-Mills theory as opposed to the standard closed string sector
that the 't Hooft expansion leads us to expect  \cite{Vijay,Holey}.   It would be useful
to understand in this case how and whether the computation of
holomorphic data in such a theory localizes to sums of Matrix
integrals.
\end{enumerate} 

In addition to these directions there are some interesting
applications that arise from known facts about the large-$N$ of the
standard double-trace $U(N)$ Matrix model.  This theory is related to
two-dimensional gravity with a positive cosmological constant and displays phase
transitions between branched polymer and smooth phases of two-dimensional gravity
\cite{Das}.   Presumably such phase transitions manifest themselves as
interesting phenomena in a four-dimensional field theory.

\section*{Acknowledgments}
Work on this project at the University of Pennsylvania was supported
by the DOE under cooperative research agreement DE-FG02-95ER40893, and
by an NSF Focused Research Grant DMS0139799 for ``The Geometry of
Superstrings".   This research was also supported by the Institute for
Advanced Study, 
under the NSF grant PHY-0070928 and the Virginia Polytechnic Institute
and State 
University, under the DOE grant DE-FG05-92ER40709.   We gratefully
acknowledge
J.~Erlich for participation in the earlier stages of the work and
D.~Berenstein, P.~Berglund, F.~Cachazo, R.~Dijkgraaf, A.~Hanany,
K.~Intriligator, 
P.~Kraus, R.~Leigh, 
M.~Mari\~{n}o, D.~Minic, J.~McGreevy, N.~Seiberg, and C.~Vafa for
comments and
revelations.  We thank Ravi Nicholas Balasubramanian for inspirational
babbling. 
BF and VJ also express their appreciation of the most generous
hospitality of the High Energy Group at the University of
Pennsylvania; they and YHH further toast to
W.~Buchanan of La Reserve B \& B for his warm congeniality.

\bibliographystyle{JHEP}

\end{document}